\documentclass[11pt, a4paper]{article}
\pdfoutput=1
\usepackage[utf8]{inputenc}
\usepackage{amsthm, amsfonts, amsmath, amssymb}
\usepackage{jheppub}
\usepackage{graphicx}
\usepackage{color}
\usepackage{siunitx}
\usepackage{hyperref}
\usepackage{cleveref}
\usepackage{commath}
\usepackage{tikz}
\usepackage{cancel}
\usepackage{comment}
\usepackage{multirow}
\usepackage[boxed, vlined, linesnumbered]{algorithm2e}

\newcommand{\be}{\begin{equation}}
\newcommand{\ee}{\end{equation}}

\newcommand{\nn}{\nonumber}
\DeclareMathOperator{\Tr}{Tr}

\title{Nonparametric Density Estimation \\from Markov Chains}

\author[a,b,c]{Andrea De Simone,}
\author[a,b]{Alessandro Morandini}

\affiliation[a]{SISSA, via Bonomea 265, 34136 Trieste, Italy}
\affiliation[b]{INFN Sezione di Trieste, via Bonomea 265, 34136 Trieste, Italy}
\affiliation[c]{School of Science and Technology, University of Camerino, Italy}

\emailAdd{andrea.desimone@sissa.it}
\emailAdd{alessandro.morandini@sissa.it}

\abstract{
We introduce a new nonparametric density estimator inspired by Markov Chains, 
and generalizing the well-known Kernel Density Estimator (KDE).
Our estimator presents several benefits with respect to the usual ones and can be used straightforwardly as a foundation in all density-based algorithms.
We prove the consistency of our estimator and we find it typically outperforms KDE in situations of large sample size and high dimensionality. 
We also employ our density estimator to build a local outlier detector, 
showing very promising results when applied to some realistic datasets.
}

\begin{document}

\maketitle

\section{Introduction}


Large and complex datasets are now commonly available in many fields such as pure sciences, medicine, engineering and finance.  The need to extract meaningful information out of them (signals, patterns, clusters, etc.) is more urgent than ever, 
and the techniques of modern statistics and data science are surging.

Machine learning (ML) prominently encompasses tools to address most data analysis tasks, e.g. outlier detection \cite{hodge2004survey, outlier_review}, classification \cite{real2019regularized,classifier_review}, clustering \cite{jain1999data,clustering_review}. For a large enough sample, there will be a complex enough architecture able to 
accomplish a given task with sufficiently high precision.
A higher degree of complexity (e.g. deep neural networks) typically corresponds to more accurate final results, but also to longer evaluation times and harder-to-interpret data processing. The latter issue  is known as the ``black-box'' problem of ML: we feed the algorithm with some input data and get some satisfactory output, but with little or no understanding of what led the algorithm to make its choices \cite{blackbox}.

Instead of focusing on complex architectures, we follow a different route and consider more traditional density-based techniques. 
A density-based approach follows a transparent course:  probability densities are estimated from the data and then used 
to infer properties of the underlying data-generating process. 

Our interest in this paper will be on the first step: we introduce here a new idea on how to estimate the probability density function of the observed data. One of the most popular approaches to density estimation is Kernel Density Estimation (KDE), which is fast, reliable and easy to understand (see e.g.~Ref.~\cite{scott2015multivariate} for a comprehensive review). In the following we propose a density estimator based on Markov Chains, and we show that it is indeed a generalization of KDE. 
Markov processes are widely applied in statistical physics as well as data analysis \cite{gilks1995markov, langville2006updating}.  It is then natural, given the strong link between Markov chains and underlying probabilities, to investigate how such a link can be used to build a density estimator.

The remainder of this paper is organized as follows.
In \cref{sec:algo} we describe our algorithm and show with numerical simulations that the estimator is consistent and 
displays good performances.
An explicit application to the task of outlier detection is described in \cref{sec:out}, where we show that we can tackle the problem in a way that is competitive with established approaches. We conclude and lay down plans for future improvements in \cref{sec:concl}.
The appendices contain the mathematical proof that our estimator is consistent.
The error analysis for two variants of our estimator is performed in \cref{sec:MSE}, with more technical details in \cref{sec:technical}.

\section{Algorithm for Density Estimation}
\label{sec:algo}

\subsection{Construction of the Markov chain}\label{sec:MC_construction}

Consider the sample $\{\boldsymbol{x}_i : \boldsymbol{x}_i\in \mathbb{R}^D\}_{i=1}^{N}$
of $N$  independent and identically distributed 
 realizations of a $D$-dimensional random vector $X$
with unknown probability density function (PDF) $f(\boldsymbol{x})$:
\be
\mathcal{S}\equiv \{\boldsymbol{x}_i\}_{i=1}^{N} 
\stackrel{\textrm{iid}}{\sim} 
f\,.
\ee
Throughout the paper we assume $N>2$.
Let $d(\cdot, \cdot)$ be a metric on $\mathbb{R}^D$, e.g. Euclidean.
Let us also consider a stochastic process starting from any $\boldsymbol{x}_j\in \mathcal{S}$ and constrained to take values
on $\mathcal{S}$, with defined
probabilities to move from one point to another.
This process can be described as a
discrete-time Markov Chain (MC) having the finite set $\mathcal{S}$ as state space.
The basic idea behind this paper is to build a density estimator out of a MC over the data sample which is spending more time (on average) in regions of higher density of points. 
We now turn to discuss how such a MC can be constructed.

In order to define the transition probabilities of the MC, we first construct the $N\times N$ distance matrix $d_{mn}$,  defined for each pair  of points $\boldsymbol{x}_m,\boldsymbol{x}_n\in \mathcal{S}$ by $d_{mn}\equiv d(\boldsymbol{x}_m,\boldsymbol{x}_n)$.
Then, we define
the $N\times N$ symmetric matrix of weights $W_{mn}$, as a function  $g$ of the distance between points
$W_{mn} \equiv g(d_{mn})\,(1-b\, \delta_{mn})$,
where $\delta_{ij}$ is the Kronecker delta and the real number $b\in [0,1]$ is what we call the ``movement bias''. In order to clarify what we mean by that, remember that $W_{mn}$ is closely related to the probability of the MC to go from $\boldsymbol{x}_m$ to $\boldsymbol{x}_n$. In this sense, the diagonal elements $W_{nn}$ describe the probability that the Markov Chain does not move. A movement bias $b=0$ corresponds to maximizing the probability that the MC does not move, while $b=1$ corresponds to forcing the chain to always move to a different point.

Some properties are required for these functions $g$: they need to be monotonically decreasing, continuous, differentiable, with bounded first derivative. As it will be clear in the next paragraphs, this $g(d)$ function is related to the kernel function $K(u)$ of KDE. 
Finally, the $N\times N$ transition matrix $Q_{mn}$, which provides the probability to move from  $\boldsymbol{x}_m$
to $\boldsymbol{x}_n$, is  obtained by normalizing the weights in $W_{mn}$ to sum to 1 on each row, as
\be
Q_{mn}\equiv \frac{W_{mn}}{\sum_{k=1}^N W_{mk}}=
\frac{g(d_{mn})(1-b\, \delta_{mn})}{\sum_{k=1}^N g(d_{mk})(1-b\, \delta_{mk})}\,,
\label{eq:transitionprob}
\ee
which in general is not symmetric because of the row-dependent
normalization. The transition probability matrix defined by \cref{eq:transitionprob} is now a right stochastic matrix with the following properties:
(1) all entries are non-negative ($Q_{mn}\geq 0$);
(2) each row sums to 1 ($\sum_{n=1}^N Q_{mn}=1$).
Since $g(d)$ is a decreasing function, it is more likely to move to closer points, according to the distance metric $d$.

Since all states in $\mathcal{S}$ communicate (i.e. $Q_{mn}>0$ for $m\neq n$), there is only one communication class and the MC is then irreducible. Since the state space $\mathcal{S}$ is finite,
the irreducible MC is also positive recurrent and thus it has
a unique stationary distribution $\boldsymbol{\pi}$ 
such that $\boldsymbol{\pi}$ is invariant under $Q$, i.e. $\boldsymbol{\pi} Q = \boldsymbol{\pi}$.
So, $\boldsymbol{\pi}$ is a left eigenvector of $Q$ with eigenvalue 1, or equivalently, $\boldsymbol{\pi}$ is 
the principal right eigenvector of $Q^T$
(Perron-Frobenius eigenvector) \cite{ross1996stochastic}.
Since we start from the weight matrix $W_{mn}$, it is trivial to derive a left eigenvector of $Q_{mn}$. As a matter of fact by using \cref{eq:transitionprob} it is easy to check that $\pi_m=\sum_{n=1}^N{W_{mn}}$ satisfies the eigenvector equation $\boldsymbol{\pi} Q = \boldsymbol{\pi}$. 

From the Markov Chain point of view, the stationary distribution $\boldsymbol{\pi}$ can be viewed as a mapping from the state space $\mathcal{S}$ of the MC to $\mathbb{R}^N$: 
for a given data sample $\mathcal{S}$, it provides an $N$-dimensional row vector whose entries sum to 1.
The stationary distribution encodes the information on the proportion of time the MC spends on a given point, in the long run:
i.e. the component $\pi_m$ is the probability to find the chain at point $\boldsymbol{x}_m$ after infinite iterations.
The density estimator we propose in this paper is built upon the observation that 
the MC constructed above is spending more time in higher density regions.
Therefore, our estimate for the probability density at point $\boldsymbol{x}_j$ is 
directly related to the $m$-th component of the principal eigenvector  
\be
\hat f(\boldsymbol{x}_m)\propto 
\pi_m\,,
\label{eq:phat}
\ee
as we are turning to describe next.
Notice that the $\boldsymbol{\pi}$  vector can be seen as the output of a map from the sample $\mathcal{S}$ of size $N$ to a vector of probabilities in $\mathbb{R}^N$.

\subsection{Connection with Kernel Density Estimation}

It is easy to see, although not obvious \textit{a priori}, that the stationary distribution of a Markov Chain with movement bias $b=0$ is proportional to the KDE estimate of the probability. Indeed, in this case we have $\pi_m\propto \sum_{n=1}^N g(d_{mn})= \sum_{n=1}^N K(d(\boldsymbol{x}_m,\boldsymbol{x}_n)/h)$, where $K(u)$ are the usual kernel functions of KDE and $h$ is the bandwidth parameter, which in this discussion is supposed to be fixed. From now on we identify our function $g(d)$ with the kernel function $K(d/h)$ and we write the stationary distribution of the MC as $\boldsymbol{\pi}^{(h)}$ wherever it is necessary to make explicit the dependence on the bandwidth parameter $h$. 
This means that the KDE estimate can be recovered from our density estimator (up to an overall factor) by setting the parameter $b=0$. However, independently of the value of $b$ it is always possible to establish a link between our density estimator and KDE. 
We remind that the KDE estimate in a generic point $\boldsymbol{x}\in\mathbb{R}^D$ is
\be 
\hat f_\text{KDE}(\boldsymbol{x})=\frac{1}{Nh^D}
\sum_{n=1}^N  K\left(\frac{d(\boldsymbol{x},\boldsymbol{x}_n)}{h}\right) \,,
\ee
where $h$ is the bandwidth parameter. Notice that a more general expression can be written by introducing multiple bandwidths parameters 
and organizing them into a $D\times D$ bandwidth matrix $H$. The effect of multiple bandwidths can be recovered by choosing a distance metric, other than Euclidean,  assigning different weights to different directions. Furthermore, after the data are pre-processed to have zero mean vector and identity covariance matrix (see \cref{sec:exp_de}), a single bandwidth parameter is enough also for multi-dimensional density estimation.
We will not pursue the case of a generic bandwidth matrix further, except that in \cref{eq:AMSED}.

For  a generic $b\in [0,1]$, the $m$-th component of the MC stationary distribution $\boldsymbol{\pi}^{(h)}$ is
\be\label{eq:kde_linkD}
\pi_m^{(h)} = \frac{1}{N}\sum_{n=1}^N W_{mn}\propto \frac{1}{Nh^D} \left[\sum_{n=1}^N K\left(\frac{d(\boldsymbol{x}_m,\boldsymbol{x}_n)}{h}\right)(1 - b \delta_{mn})\right]=\hat f_\text{KDE}(\boldsymbol{x}_m) - b\frac{\, K(0)}{N h^D}.
\ee
As anticipated in \cref{eq:phat}, our PDF estimate is proportional to  $\pi_m^{(h)}$, which is now related to the KDE estimate. By taking $b=0$ or $1$ we could recover the usual KDE estimator and the leave-one-out density estimator on the points of the sample, respectively, provided we fix the proportionality constant according to their standard definition. However, as explained in the next section, the way we fix the proportionality constant does not lead to the same final results as the estimators mentioned above. 
It is important to remark that fixing the constant in this way leads to lowering the bias in the main population with respect to KDE.


We use this link with KDE notably in two cases. First, the relation with KDE will be employed  to discuss the error analysis in \cref{sec:MSE}, since we will use known theoretical results of KDE. 
Our numerical implementation for the estimator and the optimization routines will benefit from
existing software already available for KDE 
\cite{Pedregosa:2012toh}.

\subsection{Density estimate}
\label{sec:densityestimate}

As mentioned in the previous section, for $b=0$ our estimator  recovers  KDE, where the correct PDF normalization is automatic and it is possible to evaluate the probability density in every point $\boldsymbol{x}\in \mathbb{R}^D$. 
However, in general we need to specify how to build a probability density estimate, starting from our initial estimate defined
only over the $N$ points of the sample: 
the $N$-dimensional vector $\boldsymbol{\pi}^{(h)}$ of the MC stationary distribution discussed in \cref{sec:MC_construction}. 
We also need a procedure to select the free parameter $h$ (bandwidth parameter), whose role is to allow the estimator to adapt to very different situations, while still retaining its nonparametric nature. 

We proceed in three steps: 
(1) we extend the domain of $\boldsymbol{\pi}^{(h)}$  by constructing a continuous function $q_h(\boldsymbol{x})$ defined over the whole target space of the random vector $X$, typically $\mathbb{R}^D$ for continuous random variables;
(2) we normalize $q_h(\boldsymbol{x})$ to be a probability density;
(3) we optimize over the parameter $h$.

\begin{enumerate}
\item Domain extension.

 This extension can be carried out in several ways: first of all we can perform an interpolation of the values of $\boldsymbol{\pi}^{(h)}$, but we can also take advantage of the KDE estimate and extend it to points not belonging to the sample. We have discussed an extension of the KDE estimate and the linear interpolation in \cref{sec:interp1,sec:interp2} respectively. Regardless of the interpolation procedure, some error would still be present and larger $N$ implies more accurate interpolation. Due to the fact that data are sparser at higher dimensions, it is more difficult to correctly interpolate for large $D$ and the resulting error is larger.

In the following we focus on the case where the function $q_h(\boldsymbol{x})$ is obtained through interpolation. There are several reasons why we focus on this. First of all, this is the estimator we use in practice for our numerical experiments. Secondly, it is possible to optimize this estimator in a way completely independent from KDE as we turn to describe now. Indeed, deriving $q_h(\boldsymbol{x})$ from an interpolation starting from $\boldsymbol{\pi}^{(h)}$ is the most agnostic thing we can do and the only option we had if we did not realize there was a link with KDE. An estimator that is constructed relying on an extension of KDE is presented in \cref{sec:technical}. 

\item Normalization.

The function we have now, $q_h(\boldsymbol{x})$, is defined on the correct domain, but it is not properly normalized. 
We define the new function $\tilde q_h(\boldsymbol{x})$ simply by
\begin{equation}
\tilde q_h(\boldsymbol{x}) \equiv \dfrac{q_h (\boldsymbol{x})}{\int\, q_h(\boldsymbol{x})\, \dif \boldsymbol{x}} 
\end{equation}
which integrates to 1 over the domain of $\boldsymbol{x}$.

The integration $\int q_h(\boldsymbol{x})\dif \boldsymbol{x}$ 
can be carried out in several ways, but at higher dimensions the Monte-Carlo method turns out to be effective. An integration error is still present, and it is smaller the more points are used for the Monte-Carlo, but it is larger for higher dimensionalities. The errors coming from interpolation and integration might seem like a huge problem, but for some practical applications, e.g. the outlier detector constructed in \ref{sec:exp_out}, these types of error are not relevant.

\newpage

\item Optimization. 

We have now a set of probability estimates $\tilde q_h(\boldsymbol x)$
depending on the continuous parameter $h$.
Our approach to optimization over $h$ is to minimize a loss function constructed only with our estimates $\tilde q_h(\boldsymbol{x})$. A simple  loss function we can build for this purpose is the negative log-likelihood
\begin{equation}\label{eq:optimization}
    \text{Loss}(h)=-\sum_{j=1}^N \log \tilde q_h(\boldsymbol{x}_j) \quad\longrightarrow\quad h^*=\arg\min_{h}\, \text{Loss}(h)\,,
\end{equation}
where $h^*$ is the optimal value for the free parameter.
The intuition behind this optimization is that among all the possible normalized PDFs, the ones closer to the correct $f(\boldsymbol{x})$  maximize the probability in the drawn points. This does not work for KDE, since the loss function has a global minimum for $h=0$.

We further discuss the different ways to perform bandwidth optimization in \cref{sec:tips}; if the sample is large enough a good way to optimize is use results from KDE.

\end{enumerate}
Finally, our estimate of the PDF is given by 
\be
\hat f(\boldsymbol{x}) = \tilde q_{h^*}(\boldsymbol{x})\,.
\ee
A summary of our full procedure to find the PDF estimate $\hat f(\boldsymbol{x})$ is presented in algorithm \ref{alg:tot}. 
The code for the Markov Chain Density Estimator (MCDE) with an example of usage is publicly available\footnote{
\href{https://github.com/de-simone/MarkovChainDensityEstimator}{https://github.com/de-simone/MarkovChainDensityEstimator}}.
It is up to the user to define a range of values $[h_{\min}, h_{\max}]$  where to look for the optimal value of $h$. 
At step 4 of the algorithm the left eigenvector of $Q$ is to be found; in our implementation, we used the link with KDE, but alternative numerical methods independent of KDE are also available \cite{Stewart2000}.


\begin{algorithm}[t]
\label{alg:tot}
\SetKwInOut{Input}{Inputs}\SetKwInOut{Output}{Output}
\Input{Data sample $\mathcal{S}=\{\boldsymbol{x}_i : \boldsymbol{x}_i\in \mathbb{R}^D\}_{i=1}^{N}$, \\
range of values for the bandiwdth parameter $h$: $h\,\in\,[h_{\min},\,h_{\max}]$}
\Output{Probability density estimate $\hat f(\boldsymbol{x})$ defined on $\mathbb{R}^D$}
\BlankLine
choose a family of kernel functions $K$\;

$h \leftarrow h_{\min}$\;

\While{$h \leq h_{\max}$}
{


$\boldsymbol{\pi}^{(h)}\, \leftarrow$ \cref{eq:kde_linkD} with $b=1$\;

$q_h(\boldsymbol{x})\,\leftarrow$ interpolate $\boldsymbol{\pi}^{(h)}$ over $\mathbb{R}^D$\;

$\tilde q_h(\boldsymbol{x})\,\leftarrow$ normalize $q_h(\boldsymbol{x})$   
(s.t. $\int\, \tilde q_h (\boldsymbol{x})\, \dif \boldsymbol{x} =1$)\;

Loss($h$) $\,\leftarrow-\sum_{n=1}^N \log \tilde q_h(\boldsymbol{x}_n)$\;

increase $h$\;}

 $h^*\, \leftarrow$  $\arg\min_{h}$ Loss($h$)\;

$\hat f(\boldsymbol{x})\, \leftarrow\, \tilde q_{h^*}(\boldsymbol{x})$\;

\caption{Markov Chain Density Estimator (MCDE) with bandwidth parameter optimization and $b=1$. In line 1, $K$ needs to respect the requirements mentioned in \cref{sec:algo}. We restricted our attention to families of functions depending on a single parameter $h$ only.}
\end{algorithm}


There are many possible choices for the kernel functions $K$, but for our purposes we have found that the Gaussian Kernel $K(d/h)\propto\exp(-d^2/(2h^2))$ works well. Similarly to KDE, our final estimate does not depend sensibly on the choice of the kernel, but rather on the choice of $h$. However, it might be that for some particular distributions $f(\boldsymbol x)$ and/or tasks a different set of functions works better. For our set of functions $K(d/h)$, small values of $h$ correspond to large transition probabilities to nearby points, meaning that local variations are important in order to estimate the probability. On the contrary, large values of the parameter $h$ correspond to higher probabilities for the MC to jump to distant points, hence even far away points have an impact on the probability estimate. So the role of $h$ in our estimator is analogous to the role of the bandwidth for KDE.

We have  proven  the consistency of our estimator  in general (see the error analysis in \cref{sec:MSE}). Once we have established that our estimator is consistent, we want to see how well it works compared to other density estimators.
As a benchmark for comparison we adopt the KDE. This is justified by the fact that our estimator can be viewed as a generalization of KDE.
A numerical study of the performance of MCDE is carried out in the next section.

\subsection{Numerical performance}
\label{sec:exp_de}

From now on we will work with $b=1$ and the extension of the domain of the estimate $\boldsymbol{\pi}^{(h)}$ (defined on sample points) to  $q_h(\boldsymbol{x})$ (defined on $\mathbb{R}^D$) is carried out with interpolation. As kernel functions we consider the Gaussian family.

In this section we work with $D \leq 6$. This is a case where the errors of our estimator are better under  control. Indeed, at low-dimensionality one can perform a precise linear interpolation and we can normalize the PDF by performing a Monte Carlo integration without incurring in a large integration error.

A preliminary data pre-processing step is carried out. The initial sample $\mathcal{S}$ is transformed into $\mathcal{S}_w$, with mean vector $\mu_{\mathcal{S}_w}=0^D$ and covariance matrix  $\sigma^2_{\mathcal{S}_w\mathcal{S}_w}=I_D$. The probability estimate is then performed on this new set $\mathcal{S}_w$, so there are no off-diagonal correlations and all the directions have the same weight in the final result. This means that we can safely use the Euclidean metric when we calculate distances. We have checked that such a pre-processing improves the final result for both our estimator and KDE.

We then derive $\hat f(\boldsymbol{x})$ with our estimator following the procedure outlined in the previous subsection and summarized in algorithm \ref{alg:tot}. 

%
%
%
As a measure of the error we consider the mean square error (MSE).
The mathematical details about  the MSE for our estimator are reported in \cref{sec:MSE}. 

For the numerical experiments carried out in this section, we assess the  the performance of our estimator in a controlled situation where
we know the true PDF, and we estimate $\hat f$ using both MCDE (according to the specifications detailed above) and  KDE.

\begin{figure}[t]
    \centering
    \includegraphics[width=0.9\textwidth]{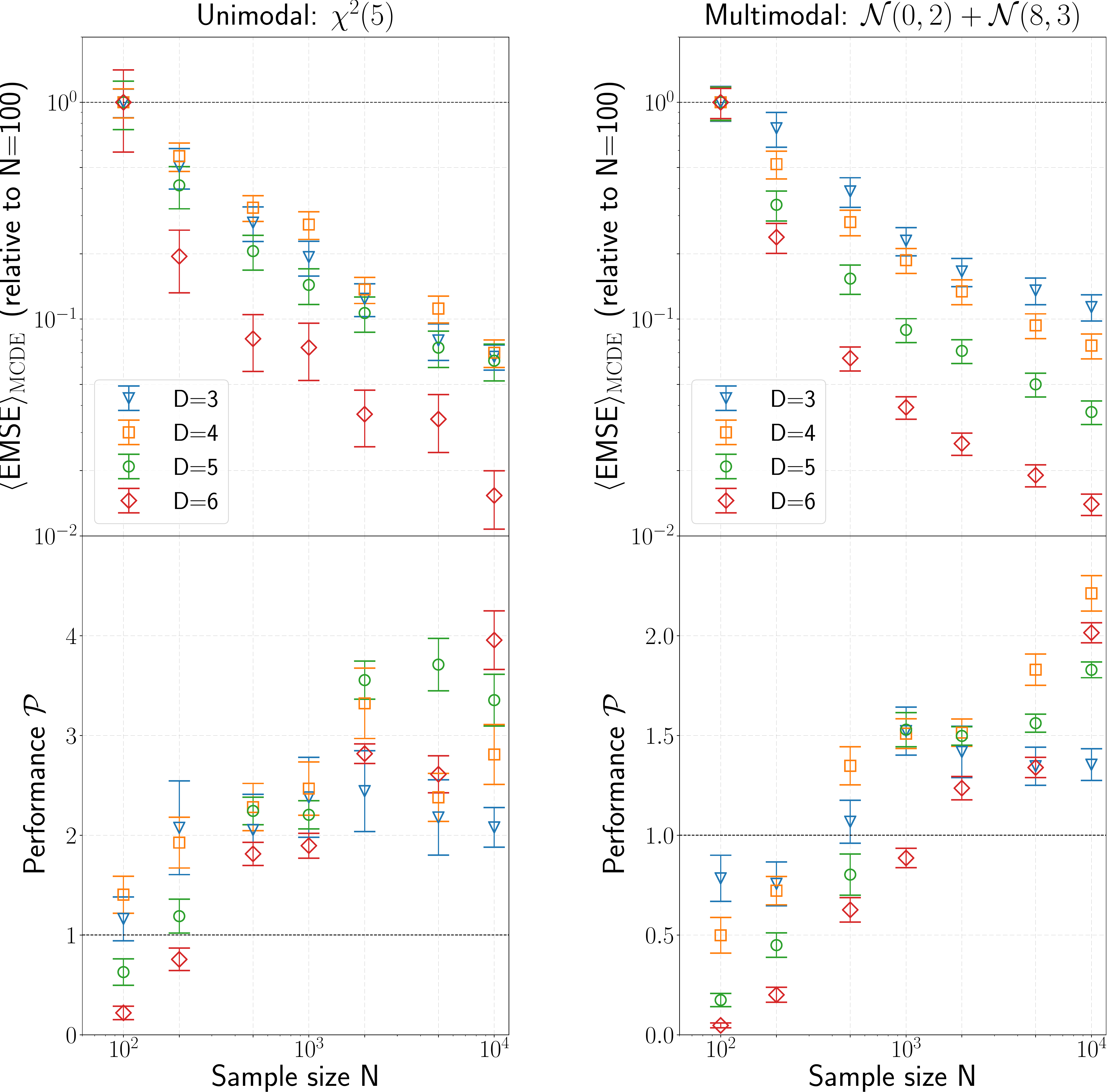}
    \caption{
    The averaged EMSE (\ref{eq:avgmse})
    and performance ratio (\ref{eq:performance}) of our estimator, for different sample sizes.
       Different bandwidth optimization have been used for the two estimators: for MCDE we used the procedure outlined in algorithm \ref{alg:tot}, for KDE we performed a 5-fold cross-validation. The error bars indicate the $2\sigma$ statistical uncertainty evaluated with $R=16$ sample realizations. The $\langle\text{EMSE}\rangle$ for a value of $N$ is scaled by dividing by the $\langle\text{EMSE}\rangle$ obtained for $N=100$. }
    \label{fig:DE}
\end{figure}

\begin{figure}[t]
    \centering
    \includegraphics[width=0.9\textwidth]{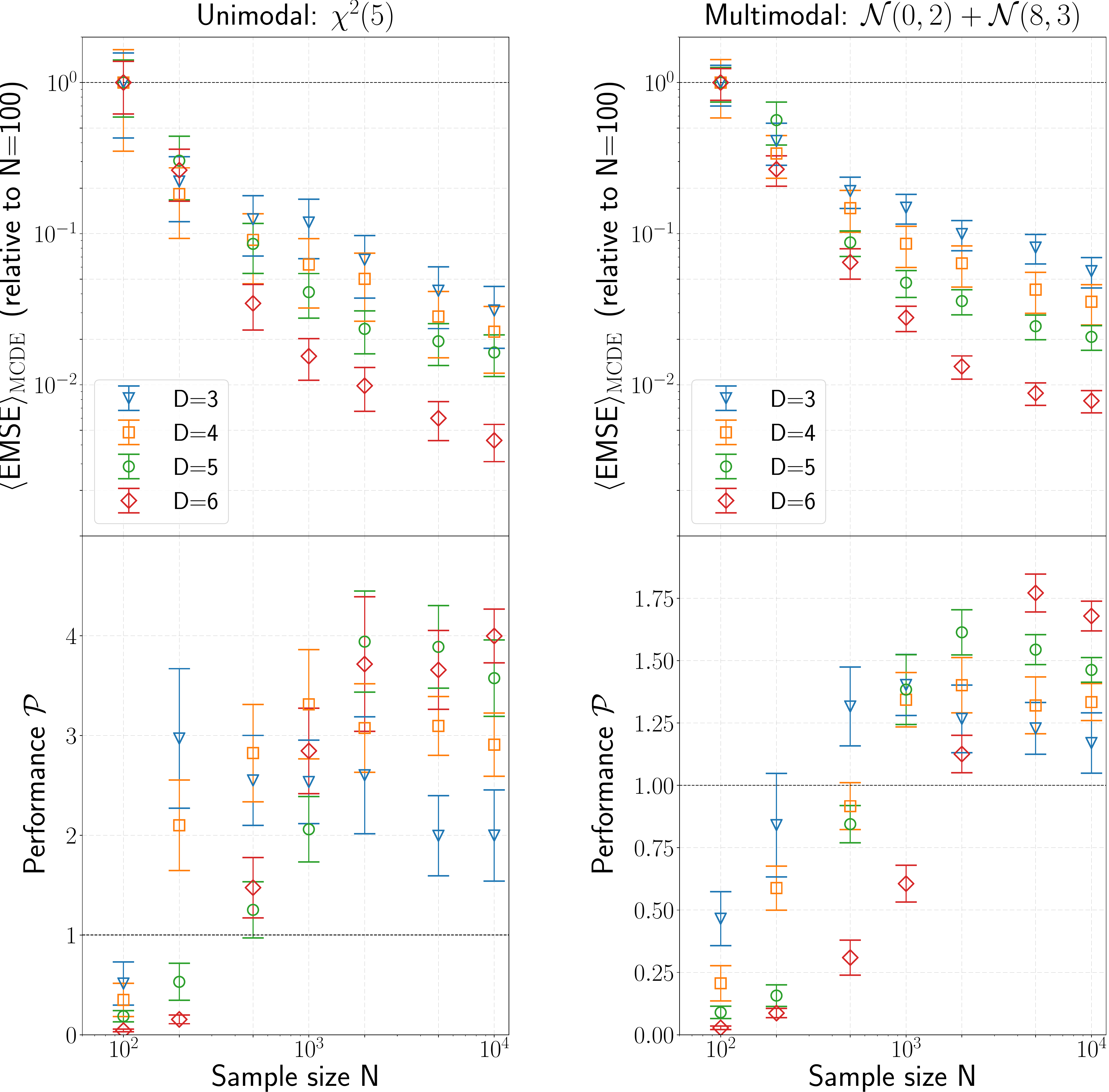}
    \caption{
    The averaged EMSE (\ref{eq:avgmse})
    and performance ratio (\ref{eq:performance}) of our estimator, for different sample sizes.
    For both MCDE and KDE the bandwidth used is the same, chosen by performing 5-fold cross-validation on KDE. The error bars indicate the $2\sigma$ statistical uncertainty evaluated with $R=9$ sample realizations. The $\langle\text{EMSE}\rangle$ for a value of $N$ is scaled by dividing by the $\langle\text{EMSE}\rangle$ obtained for $N=100$.}
    \label{fig:DE2}
\end{figure}

Given a sample $\mathcal{S}$ of size $N$, the
empirical MSE  can be computed as a sample average:
\begin{equation}
\text{EMSE}(\mathcal{S})=\frac{1}{N}\sum_{j=1}^N [f(\boldsymbol{x}_j)-\hat f(\boldsymbol{x}_j)]^2.
\end{equation}
For our estimator it is natural to average the MSE over the sample points, since we started from the vector $\boldsymbol{\pi}$ defined for each sample point. Integrating the MSE would imply giving great relevance on the way we go from $\boldsymbol{\pi}$ to $q(\boldsymbol{x})$, which is not the focus here. 
Our procedure to assess the performance of our MCDE estimator relative to KDE, in a setup where the underlying PDF is known,  is as follows:
\begin{enumerate}
    \item[(a)] PDF construction. We fix the dimensionality $D$, the number of points $N$ and the unidimensional PDF $f(x)$. For $D>1$ we multiply the PDF along each direction, so that $f(\boldsymbol{x})=f(x^1,\ldots,x^D)=f(x^1) \cdots  f(x^D)$.
    \item[(b)] Parameter selection. We generate a training set of $N$ points from the distribution and take the optimal values of bandwidth. Since we want to fix these parameters, in order to be more cautious we can do this step more than once and take the average of the values we find each time for the bandwidths. Different optimization procedures are possible here.
    \item[(c)] Testing. We generate a test set of $N$ points from the distribution.
    We estimate $f(\boldsymbol{x})$ with both MCDE and KDE, with fixed values of bandwidth found in step 2. Using fixed values for the bandwidth avoids the possible overfitting we could encounter by deriving the optimal values each time we generate a sample and makes the results easier to interpret. Then we evaluate the error we do with both KDE and MCDE. The procedure is repeated $R$ times because we want to estimate the statistical uncertainty on the errors and performances, by averaging over $R$ different realizations of the random sample.
    \item[(d)] Performance. We compute the EMSE averaged over $R$ independent realizations of the data sample, 
    \be
    \langle \text{EMSE} \rangle \equiv \frac{1}{R}\sum_i^{R} \text{EMSE}(\mathcal{S}_i)\,,
    \label{eq:avgmse}
    \ee 
    and the performance of MCDE relative to KDE, defined by the ratio
    \be 
     \mathcal{P} \equiv  \frac{ \langle \text{EMSE} \rangle_\textrm{KDE}}{ \langle \text{EMSE} \rangle_\textrm{MCDE}}\,.
     \label{eq:performance}
    \ee 
    where $\mathcal{P}>1$ means that the error of the density estimate carried out with MCDE is smaller than the one with KDE, so MCDE performs better than KDE.

\end{enumerate}
The results for the $\langle$EMSE$\rangle$ of our estimator and its performance $\mathcal{P}$ relative to KDE are shown in \cref{fig:DE,fig:DE2}, as functions of the sample size $N$, at different dimensions $D=3,4,5,6$, and for
two example cases of underlying PDF $f(x)$: a unimodal and a multimodal distribution. As a unimodal distribution (left panels of \cref{fig:DE,fig:DE2}) we choose $\chi^2$ with 5 degrees of freedom.  
As  a multimodal distribution (right panels of \cref{fig:DE,fig:DE2}) we choose a mixture of normal PDFs: $\mathcal{N}(0,2)+\mathcal{N}(8,3)$, where $\mathcal{N}(\mu,\sigma^2)$ is a normal PDF with mean $\mu$ and variance $\sigma^2$. 
Our choices are dictated by simplicity and by the fact the selected PDFs have properties relevant for our study. The $\chi^2$ distribution is peaked, but shows asymmetry around the maximum. The multimodal is asymmetric and also has several peaks, which are typically difficult for an estimator to identify.

The two \cref{fig:DE,fig:DE2} differ by the optimization procedure. 
For what concerns KDE, we scan over a range of values of bandwidth and for each value do a 5-fold cross validation with the negative log-likelihood as loss function. 
Since in the simulations we know the true $f(\boldsymbol{x})$, we are able to check explicitly whether our procedure finds optimal or near-optimal values for both MCDE and KDE. Other optimization procedures have been considered for KDE (10-fold cross-validation,  leave-one-out cross validation and the rule of thumbs by Silverman  \cite{silverman1986density} and Scott \cite{scott2015multivariate})
and lead to analogous results. We chose the 5-fold cross validation for KDE since for our simulations it gave the most stable results. 
As mentioned in \cref{sec:tips}, the KDE optimal bandwidth found by 5-fold cross validation works well also for our estimator MCDE, provided the sample is large enough. 
We show the results of numerical simulations performed with  two types of bandwidth optimization:
in \cref{fig:DE}, we use seperate optimizations for MCDE (\cref{eq:optimization}) and KDE (5-fold cross-validation); 
in \cref{fig:DE2}, the optimal $h$ for KDE is used also for MCDE.

Note that steps (b) and (c) above can be done multiple times in order to take the mean value of the optimal parameters and of the errors respectively. In our case we repeated step (b) twice and for step (c) we took $R=16$, $R=9$ for generating the plots in \cref{fig:DE,fig:DE2} respectively. 
The bandwidth values used for optimization are evenly spaced in the logarithmic interval $[10^0,10^2]/\sqrt{N}$. Dividing by $\sqrt{N}$ we ensure that the optimal value always lies within the range defined for any $N$ and any $D$.

In the plots we  show only the statistical error associated with the $R$ independent realizations of the data. We have roughly estimated the systematic errors coming from sub-optimal selection of $h$ and they are around the same order of the statistical error showed. 

The upper panels of \cref{fig:DE,fig:DE2}  show the error reduction with increasing  sample size. For a better representation we have chosen to show the relative decrease compared to a benchmark with $N=100$, so what is actually depicted is the ratio of $\langle \text{EMSE}\rangle$ for generic $N)$  divided by $\langle \text{EMSE}\rangle$ for $N=100$.
Looking at the upper row the consistency of our estimator for increasing $N$ is clear. Here the contributions to the error come from both the intrinsic error of the estimator and the normalization error. These types of error are both supposed to decrease with increasing size of the sample. 
The situations at higher $D$ are the ones that suffer the most from these error, so it is not surprising that the convergence is steeper for $D=6$ rather than for $D=3$.

 The lower panels of \cref{fig:DE,fig:DE2} show the performance with respect to KDE.
 The behavior of the curves is less regular, this is because here also the uncertainties on the $\langle \text{EMSE}\rangle$ of KDE contribute to the total error. Here systematics are of the order of twice the statistical error. Anyway, compatibly with the errors, MCDE indeed provides an enhancement of the performance relative to KDE. Here we need to distinguish again between the intrinsic errors of MCDE and the normalization error. Our estimator works better than KDE at higher dimensionality, but this gain is hidden for small samples due to the big uncertainty from normalization, whereas KDE always returns a correctly normalized function. In order for the improvement to become apparent it is necessary to reduce the error coming from the normalization and hence increase the sample size. The sparser the data, the more points are needed. For instance for the multimodal at $D=4$ having 500 points is enough to already perform better than KDE, but at $D=6$ at least 2000 points are necessary.

We can see that the two different optimization procedures give consistent results, but there are some differences. First of all, the plot of \cref{fig:DE} has a less regular behavior, especially in the lower row. This is because the performance is not only influenced by the relative performance of the estimators, but also by the goodness of the optimization procedure. We know that the optimization procedure of \cref{fig:DE2} should provide good results for our estimator only for large samples. Indeed, this can be seen in the plots. The convergence of the plot in \cref{fig:DE2} seems faster in the upper row, but this is due to the poor performance for small $N$, not to a greater improvement at large N. 

In our procedure we were able to mitigate overfitting by performing the probability estimate on samples different from the ones used for the parameter  estimation. Typically, in real situations only one sample dataset is available and it must be used for both the parameter selection and the testing. It is then natural to wonder whether our algorithm works well when the parameters are estimated from the same set used for testing. In this situation, we have explicitly checked that  we can still perform better than KDE, provided we have enough points, and that our overfitting is manageable or at least it does not compromise the performance. The case where one sample is used for both parameter optimization and testing is the one of outlier detection and we will see that our density estimator can work well for this task in \cref{sec:exp_out}.

\section{Application to Outlier Detection}\label{sec:out}

While estimating the probability distribution of data can be useful \textit{per se}, it is often just the first step of a more in depth analysis. As a matter of fact, many data analysis tasks are based on density estimation \cite{hido2011statistical,latecki2007outlier,shih}. 
In the following we focus on a density-based approach to outlier detection, where the probabilities are estimated with MCDE. Of course, our density estimator has a wider variety of applications, wherever an estimated PDFs is needed  \cite{doi:10.1002/widm.30, saito1995local}.

\subsection{Description of the method}
\label{sec:desc_out}

Given a dataset, the aim of outlier (or anomaly) detection is finding the elements which are most different from the main population of ``normal'' points. The task of finding anomalies is very important in real-world situations, for instance it is fundamental in order to detect frauds, but it has applications also in high energy physics, where beyond the standard model signals are expected to constitute a minority population \cite{Collins:2018epr}. 

Local outlier detection is typically carried by assigning an anomaly score to every point of the dataset in order to establish which points are more likely to be anomalies. The way we compute the anomaly score for any given point in a density-based approach is by comparing the density in that point with the average density in the nearby points. An outlier will most likely be distant from the main population and hence have low probability compared to its neighbors.
%
In practice, we  use an anomaly score similar to the one of \cite{tang2016local}. Given a set of points and an estimate $\hat f$ of the underlying PDF, we can assign a score $S_k(\boldsymbol{x}_i)$ to a specific point $\boldsymbol{x}_i$ by considering its $k$ nearest neighbors $\{\boldsymbol{x}_i^{(1)},\ldots,\boldsymbol{x}_i^{(k)}\}$ and averaging the density over 
them
\begin{equation}\label{eq:score_out}
    S_k(\boldsymbol{x}_i)=\dfrac{\langle \hat f(\boldsymbol{x}_i) \rangle_k}{\hat f (\boldsymbol{x}_i)}\,, \quad \text{where} \quad \langle \hat f(\boldsymbol{x}_i) \rangle_k
    \equiv\dfrac{1}{k}\sum_{l=1}^k \hat f (\boldsymbol{x}_i^{(l)})\,.
\end{equation}
The greater the score, the more likely the point is an anomaly. This score is an indicator of how much our point is declustered from nearby points. Notice that we are dealing with a local outlier detector: the anomaly score depends on the number $k$ of nearest neighbors we consider and the final performance  depends on local properties.

Since the anomaly score only depends on the values of the estimated PDF evaluated on the sample points, the errors coming from interpolation and integration are much less relevant. 
Indeed, the overall rescaling with the normalization cancels out in the ratio and we can work directly with $\hat f(\boldsymbol{x}_i)=\pi_i^{(h^*)}$. Of course we still need to find the optimal $h^*$ and, to this end, we need to work with normalized PDFs since we use \cref{eq:optimization}. However, the interpolation/integration error may only result in a sub-optimal selection of $h^*$ and this does not compromise substantially the performance of the detector, as it will be clear in the next sections. Actually, it is important to mention that this optimization procedure is the one  yielding the best result. 

One might wonder why we bother normalizing the PDFs if other (faster) optimization procedures would still provide a value for the optimal bandwidth. As explained in \cref{sec:tips} these other optimization procedures rely on KDE results and how well this optimization adapts to MCDE depends on the sample size. 
In \cref{sec:outlier_real} we work with high dimensional samples, where the optimal bandwidth for MCDE derived by minimizing the negative log-likelihood is different from the one derived with an optimization procedure relying on KDE and ultimately leads to better performance. 

There is another reason why it may be worth normalizing the PDF. In the ratio of \cref{eq:score_out} we use the same density estimator for both the numerator and the denominator. A threshold value, e.g. $S_k(\boldsymbol{x})>1$, assesses whether or not a point is to be considered as an outlier. If we use two different density estimators (one for the numerator and one for the denominator) then, in order to retain the same threshold on $S_k(\boldsymbol{x})$, we need the numerator and denominator to be comparable, hence with the same normalization. We leave further investigations of this situation to future work.

Normalizing the PDFs is useful also in a semi-supervised approach.
For instance, suppose a data sample without anomalies is available and can be used as a training set to determine a baseline PDF. In such a case, 
one can construct a local anomaly score as the ratio between the PDF estimated directly on the points of the test dataset (possibly containing anomalies) and the PDF in the same points, as derived from the training set. Averaging over the nearest neighbors helps also in this case, in order to avoid some noise. This is a direction we will pursue in the future, especially given the different weight our estimator gives to the tails of a distribution with respect to KDE.

\subsection{Numerical experiments}
\label{sec:exp_out}
\subsubsection{Synthetic datasets}

Our first round of numerical experiments is aimed at evaluating the performance of the outlier detector based on our density estimator in controlled situations, before applying it to real datasets in the next subsection.
As performance metric we use the Area Under the receiving operator characteristics Curve (AUC) for different values of $k$. 
We use linear interpolator and Monte Carlo integration for the domain extension and normalization steps of MCDE (see \cref{sec:densityestimate}), but as mentioned in   \cref{sec:desc_out} once we have fixed $h$ to its optimal value there is no need to normalize our estimated PDF.



We generate three  different sets of synthetic data, for different inlier and outlier distributions and dimensionality $D=2,\,4,\,6$.
In all the datasets, we consider the case where the outliers are localized, meaning that they are concentrated around some values described by the underlying PDF.
We have a one-dimensional inlier distribution $f_{\rm in}(x)$ and an outlier distribution $f_{\rm out}(x)$. Once we have fixed the distributions, the dimensionality and a real number $c\in[0,1]$, the  distribution employed for sampling data is given by $c\,f_{\textrm{in},D}(\boldsymbol{x})+(1-c)\,f_{\textrm{out},D}(\boldsymbol{x})$, where  $c/(1-c)$ is the imbalance ratio. 
In order for the score  in \cref{eq:score_out} to be able to localize outliers correctly, the neighborhood of each outlier point must include inlier points. 
To this end, it is sufficient (but not necessary) that the number $k$ of nearest neighbors obeys $k>(1-c)N$, where $N$ is the total sample size (inliers plus outliers).
It is also possible that the outliers are not confined within a small region, in which case this procedure works the same way, but there is not a defined minimum value for $k$. That would be for instance the case when the outliers are uniformly distributed in the domain of the problem.

\begin{figure}[t]
    \centering
    \includegraphics[width=0.5\textwidth]{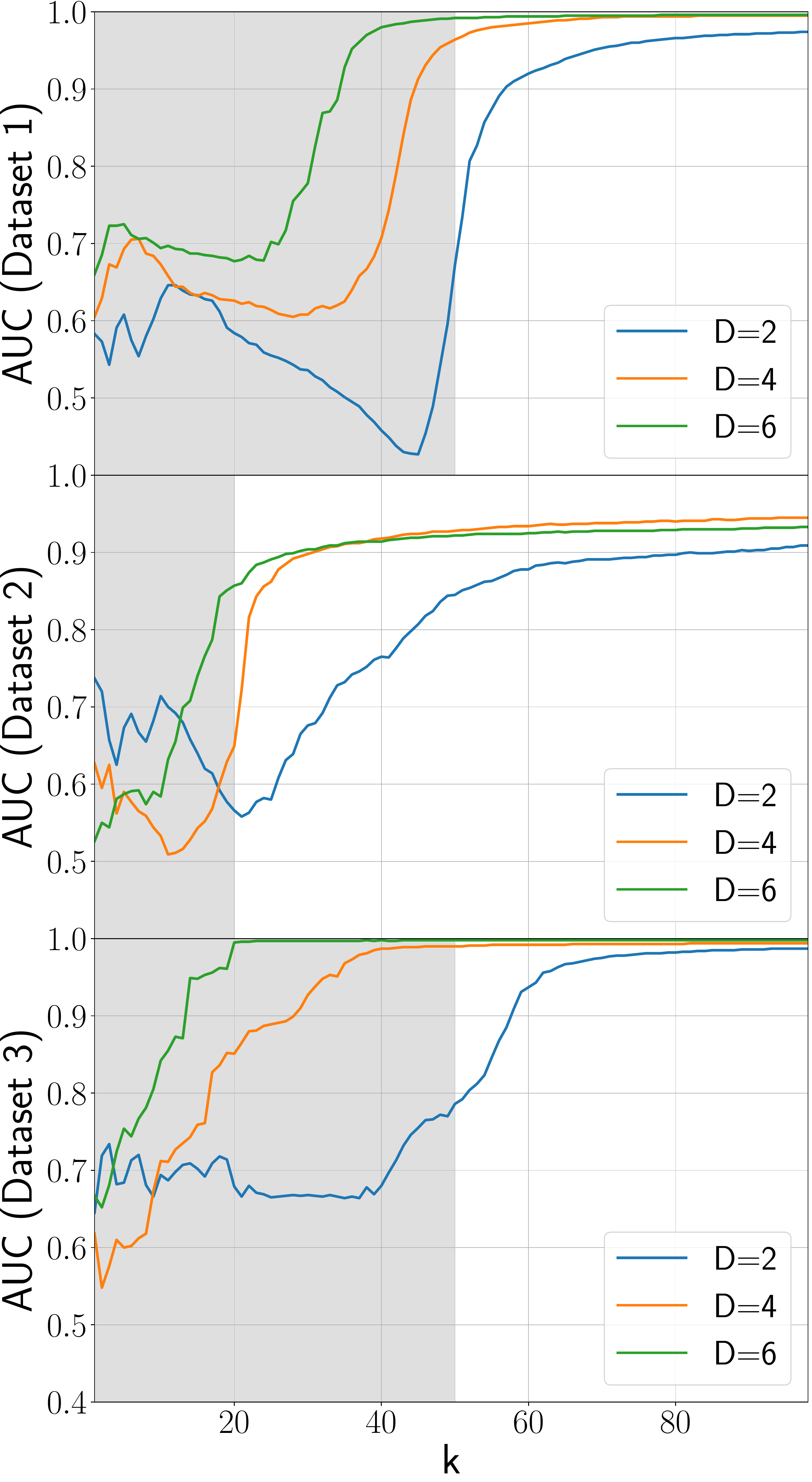}
    \caption{AUC performances for our three datasets at different dimensions, for varying number of first neighbors on the $x$ axis. A shaded grey area is present to indicate values of $k<(1-c)N$ for which we do not expect a good performance, as discussed in the text.}
    \label{fig:out_cases}
\end{figure}

The details of the three synthetic datasets used in our experiments are as follows:
\begin{enumerate}
    \item Dataset 1: $N_{\rm in}=450,\,N_{\rm out}=50$. $f_{\rm in}=\mathcal{N}(4,0.5)$, $f_{\rm out}=$ Log-Laplace$(2,1)$
    \item Dataset 2: $N_{\rm in}=180,\, N_{\rm out}=20$. $f_{\rm in}=$ Exp$(1)$, $f_{\rm out}=\mathcal{N}(5,1)$ 
    \item Dataset 3: $N_{\rm in}=950,\, N_{\rm out}=50$. $f_{\rm in}=$ Gamma$(2)$, 
    $f_{\rm out}=$ Gamma$(12)$
\end{enumerate}
The imbalance ratio is given by  $N_{\rm in}/N_{\rm out}$.
The results for the AUC are presented in \cref{fig:out_cases}. A clear trend in the dimensionality 
$D$ and the number of nearest neighbors $k$ is clear.
The shaded grey areas represent the values of $k<(1-c)N=N_{\rm out}$ for which we know the performance might not be satisfactory. If there are $N_{\rm out}$ outliers all localized in a region, 
we would need $N_{\rm out}$ neighbors in order to be able to see some inlier points, at least in the worst case scenario. It can be seen that in the grey area the performance is very irregular exactly because of this. On the other hand, for sufficiently large  $k$, the performance increases with increasing number of neighbors. This is the case only for localized outliers.

 
As the dimensions increase, the data become sparser, a signal of the curse of dimensionality.
This implies that the outliers get farther from the main inlier population and easier to identify. As a matter of fact, in higher dimensions the AUC performance is usually larger with respect to lower dimensionality (for the same value of $k$). Stated a bit differently,  
to reach the same level of performance at higher dimensions, smaller values of $k$ are needed.



\subsubsection{Real datasets}\label{sec:outlier_real}

We now turn to explore some applications of the outlier detector based on our density estimator to more realistic datasets. In most of the cases under consideration,
the data have dimensions $D>6$. 
This implies that the linear interpolation  employed previously for $D<6$  might be no longer feasible, as the computing time scales exponentially with the dimensionality. For the applications with $D>6$ we will use a nearest neighbor interpolator, which is less precise, but runs with a shorter computation time even at high dimensions. 
For the PDF normalization step, we  still use the Monte Carlo integration, though we are aware that at high dimensions the calculation will not be accurate, unless the sample size is very large. However, we will see that the performances are still competitive in comparison to other methods. The reason for this is that the bandwidth optimization still proceeds smoothly in this high-dimensional setup, as our negative log-likelihood loss function presents a clear minimum.
\begin{table}[t]
    \centering
    \begin{tabular}{|c|c|c|c|c|}
        \hline
        Dataset & Method & $k=5$ & $k=10$ & $k=20$ \\
                \hline
                \hline
     Breast cancer \cite{breastcancer}       & MCDE & 0.95 & 0.96  &  0.96\\
 $N_{\rm in}=347, \,N_{\rm out}=10$    &  RDOS & 
 \underline{0.97} & \underline{0.98} & \underline{0.98} \\
  $D=30$           &  LOF & 0.93& \underline{0.98}& \underline{0.98} \\
              \hline
  Pen-local    \cite{penlocal}       &  MCDE & \underline{0.98} & 0.98 & 0.98 \\
$ N_{\rm in}=6704,\, N_{\rm out}=10$&    RDOS & \underline{0.98} & \underline{0.99} & \underline{0.99} \\
 $D=16$        &     LOF & 0.90& \underline{0.99}& \underline{0.99}\\
              \hline
   Pen-global   \cite{penglobal}    &    MCDE & \underline{0.70} & \underline{0.86} &  0.95\\
$N_{\rm in}=629,\, N_{\rm out}=90$  &  RDOS & 0.52 & 0.73 &\underline{0.98} \\
$D=16$            &  LOF & 0.51& 0.62& 0.88 \\
              \hline
  Satellite    \cite{satellite}       & MCDE & \underline{0.79}& \underline{0.80}& \underline{0.92} \\
$N_{\rm in}=4950,\, N_{\rm out}=75$  & RDOS & 0.73 & 0.77 &0.71\\
 $D=36$           &  LOF & 0.73& 0.75& 0.68 \\
              \hline
    \end{tabular}
    \caption{The AUC performance for different datasets is presented for some values of $k$. $k$ can be interpreted as a measure of locality. The values for RDOS and LOF are taken from the plots of \cite{tang2016local}.}
    \label{tab:out_bench}
\end{table}

Outlier detection is plagued by the same issue as binary classification of imbalanced classes: the outliers (minority class members) are typically much less than the inliers (majority class members).
This makes it hard to establish a performance metric and to compare algorithms. 
Another implication is that splitting the data  into training, validation and test sets
would result into a validation/test sets with very few (if any) outliers. 
So it is difficult to find the hyperparameters  by a regular cross-validation procedure. 
Of course we can still fix $h$ for MCDE as we have done before, but in a purely agnostic approach we have no way of fixing other parameters, e.g. the number $k$ of nearest neighbors.

The work in Ref.~\cite{tang2016local} was developed in this spirit. In this work there is a comparison between different algorithms treating $k$ as a free parameter, 
ultimately considered as a measure of outlier locality. We compare their results with ours in \cref{tab:out_bench}. Here we consider only the two algorithms which work best, namely RDOS \cite{tang2016local} and LOF \cite{lof}. The value of $k$ is not fixed, so it is important to compare the algorithms for different values. In Ref.~\cite{tang2016local}, values of $k$ between 1 and 32 were considered, so we focused our attention on three benchmark values $k=5,10,20$. 

The results of \cref{tab:out_bench} show that our outlier detector is competitive with the others in the first two datasets, being at most $2\%$ below the best performer. 
For the last two datasets our outlier detector either perform comparably or considerably better than the others, with up to a $30\%$ improvement over the second best result. For all datasets our estimator works well for $k=5$, indicating that our algorithm correctly captures local deviations.

\begin{table}[t]
    \centering
    \begin{tabular}{|c|c||c|c|c||c|c|c|}
         \hline
        \multicolumn{1}{|c}{\multirow{2}{*}{Dataset}} 
        & \multicolumn{1}{|c||}{\multirow{2}{*}{Method}}
        & \multicolumn{3}{c||}{\small $k=\sqrt{N}$} & \multicolumn{3}{c|}{\small $k=3\sqrt{N}$}\\
        \cline{3-8}
        \multicolumn{1}{|c}{}
        &\multicolumn{1}{|c||}{} & $c=.9$  & $c=.95$ & $c=.98$ & $c=.9$  & $c=.95$ & $c=.98$\\
                \hline
                \hline
                     Ionosphere \cite{UCI,iono} & MCDE & \underline{0.93} &  \underline{0.95} & \underline{0.98} & 0.93 &  \underline{0.95}  & \underline{0.98}\\
     $N_{\rm in}=225\,,D=34$ & LOF & 0.91 & 0.92  & \underline{0.98}  & \underline{0.94} &  \underline{0.95} & 0.97\\
      \hline
     Heart \cite{UCI} & MCDE & 0.63 & \underline{0.74} & \underline{0.84} & 0.70 & \underline{0.74} & 0.82\\
     $N_{\rm in}=150\,,D=13$ & LOF & \underline{0.69} & \underline{0.74} & 0.80 & \underline{0.73} & 0.73 & \underline{0.84}\\
      \hline
           Iris \cite{UCI}& MCDE & 0.74 & 0.84 & 0.78 & 0.83 & \underline{0.86} & \underline{0.78}\\
     $N_{\rm in}=100\,,D=4$ & LOF & \underline{0.85} & \underline{0.89} & \underline{0.83} & \underline{0.88} & 0.82 & 0.68\\
      \hline
           Parkinson \cite{parkinson} & MCDE  & \underline{0.42} & 0.50 & 0.58  & 0.43 & 0.51 & \underline{0.57}\\
      $N_{\rm in}=147\,,D=22$& LOF  & \underline{0.42} & \underline{0.52} & \underline{0.62} & \underline{0.44} & \underline{0.52} & 0.56\\
      \hline
                 Transfusion \cite{transfusion}& MCDE  & \underline{0.61} & \underline{0.74} & \underline{0.77} & 0.61 & 0.73 & \underline{0.74}\\
     $N_{\rm in}=570\,,D=4$ & LOF  & 0.60 & 0.70 & 0.76 & \underline{0.69} & \underline{0.76} & 0.73\\
      \hline
           Vehicle \cite{UCI,vehicle}& MCDE  & 0.78 & 0.85 & 0.87 & \underline{0.77} &\underline{0.83} & \underline{0.85}\\
      $N_{\rm in}=647\,,D=18$& LOF  &\underline{0.85} & \underline{0.90} & \underline{0.91} & 0.76 & 0.81 & \underline{0.85}\\
      \hline
                 CMC \cite{UCI}& MCDE  & 0.43 & 0.48 & \underline{0.46} & \underline{0.40} & \underline{0.45} & \underline{0.43}\\
     $N_{\rm in}=1140\,,D=9$ & LOF  & \underline{0.45} & \underline{0.49} & \underline{0.46} & \underline{0.40} & \underline{0.45} & \underline{0.43}\\
      \hline
                 Yeast \cite{UCI, yeast} & MCDE & \underline{0.47} & \underline{0.48} & \underline{0.53} & 0.48 & 0.51 & 0.56\\
     $N_{\rm in}=1240\,,D=8$ & LOF  & 0.46 & 0.47 & 0.50 & \underline{0.51} & \underline{0.56} & \underline{0.58}\\
      \hline
           PC1 \cite{pc1}& MCDE  & - & \underline{0.70} & \underline{0.72} & - & \underline{0.71} & \underline{0.72}\\
      $N_{\rm in}=1032\,,D=21$& LOF  & - & 0.67 & 0.70 & - & 0.69 & 0.69\\
      \hline
      \end{tabular}
    \caption{AUC for eight benchmark datasets. We have considered two reasonable choices of $k$ and three different imbalance regimes ($c=N_\text{in}/N$).}
    \label{tab:out_bench2}
\end{table}

Of course the performance of the outlier detector depends on many factors, but the most relevant one is the imbalance ratio, i.e. the ratio of the numbers of inliers and outliers. 
We then proceed to consider some benchmark datasets known in the literature, used for instance in \cite{gensample}, and explore the consequences of changing the number of outliers. 
For comparison we have used LOF, as it was easy to implement and based on a free parameter $k$, so that the comparison with our algorithm is meaningful. 
%
%
The results are presented in \cref{tab:out_bench2}, where we vary the imbalance parameter $c=N_{\rm in}/N$ to have values 0.9, 0.95, 0.98. 
As for the values of $k$, the usual rule-of-thumb of nearest neighbors is $k=\sqrt{N}$. 
However, from our previous analysis of the synthetic datasets we also know that 
large values of $k$ should saturate the performance, if we have some reason to believe the outliers are localized.
Therefore, we also report the results for $k=3\sqrt{N}$. 
These results indicate that our outlier detector performs better for larger imbalance ratios, 
i.e. when there are very few outliers.

It is well known that it is not possible to have an algorithm that performs best in any possible setup \cite{wolpert1997no}. 
As mentioned in the introduction, our algorithm is not meant to outperform the  state-of-the-art methods for density estimation or outlier detection, especially those based on deep neural networks. Our goal in this section was to show that an outlier detector constructed straightforwardly out of our MCDE is competitive with (and, in many cases, better than) other easy-to-interpret approaches.

\section{Conclusions and outlook}\label{sec:concl}

In this paper we presented the Markov Chain Density Estimator (MCDE), 
a novel nonparametric density estimator based upon Markov Chains.
MCDE can be seen as a generalization of the Kernel Density Estimator (KDE).
We proved the consistency of MCDE and showed it in 
practice with numerical experiments.
We also carried out a comparative analysis of the performance of MCDE with respect to KDE
and concluded MCDE works better at higher dimensions and for sufficiently large sample size.

In addition to be a highly performant density estimator in itself,
MCDE can be used as a baseline tool for several other tasks based on probability densities.
As an application of our density estimator we considered local density-based outlier detection, both on synthetic and real datasets, and highlighted the role of outlier locality.
Indeed, the outlier detector built out of MCDE performed very well on synthetic datasets and
better than other density-based approaches on real datasets.



Several directions for further investigations of MCDE can be envisaged, both on the computational side and on the application side.
The main uncertainties affecting the density estimate come from interpolation and integration errors. The use of  more sophisticated interpolators/integrators than the ones employed in this paper may improve the performance of our estimator even further.
Furthermore, in conjunction with dimensionality reduction techniques, 
such as Principal Component Analysis or Variational Auto-Encoders, 
our estimator can prove very useful for outlier detection in high-dimensional datasets.
An accurate estimation of the probability density underlying a data sample may also be used as a generative model or for clustering purposes.

\section*{Acknowledgements}

We would like to thank Alessandro Davoli for insightful discussions in the preliminary phases of this project.

\appendix

\section{Error analysis}\label{sec:MSE}

In this appendix we will show the conditions under which our estimator is consistent. We remind that for KDE the necessary requirements in $D=1$ are that as $N\rightarrow \infty$, we also have that $h\rightarrow 0$ and $Nh\rightarrow \infty$. In our case a further requirement needs to be satisfied as we also need $V(N)/Nh\rightarrow 0$, where $V(N)$ is the volume of the convex hull of the sample points. This latter condition can be satisfied as long as $V(N)$ goes to infinity slower than $N$, a condition satisfied if for instance $f(x)$ admits finite first and second moments.

In order to prove the consistency of our estimator it is useful to relate it to KDE, so that known results can be used and the comparison between KDE and MCDE will be easier. We  carry out the error analysis in the 1D case in detail, and at the end of this section we comment on how the generalization at higher $D$ proceeds.

After the eigenvector equation we know that our pointwise estimator can be written as
(see \cref{eq:kde_linkD})
\be\label{eq:interp0}
\hat f(x_i)\propto \pi_i^{(h)} \propto \frac{1}{N}\sum_{k=1}^N W_{ik}= \frac{1}{N h} \left[\sum_{k=1}^N K\left(\frac{x_i-x_k}{h}\right)- b K(0)\right]=\hat f_\text{KDE}(x_i) - b\frac{\, K(0)}{N h},
\ee
where $K(x)$ is a Kernel function such that $\int K(x)\, \dif x=1$
and $\int x\,K(x)\, \dif x=0$. This is a pointwise estimator and we want to extend its domain to the real axis. We will do it in two different ways and get two different density estimators: $\hat f_1(x)$ and $\hat f_2(x)$. First we consider extending the KDE estimate to points not belonging to the sample
\be\label{eq:interp1}
\hat f_1(x)=C_1\left(\hat f_\text{KDE}(x) - b\frac{\, K(0)}{N h}\right)\, I_{\mathbb{D}_1}(x), \quad \text{ with }
\mathbb{D}_1=
\left\{x \left\vert\hat f_\text{KDE}(x)\geq b\frac{K(0)}{Nh}\right.\right\},
\ee
where $I(x)$ is the indicator function and $\mathbb{D}_1$ is the part of the real axis where $\hat f_\text{KDE}(x)- b K(0)/N h\geq0$. Notice that $\mathbb{D}_1$ is not compact in general.  The constant $C_1$ is to be fixed by requiring the normalization. With these definitions,  $\hat f_1(x)\geq 0, \forall x\in \mathbb{R}$ and $\int_\mathbb{R} \, \hat f_1 (x)\, \dif x = 1$.

The second estimator we build is constructed on a linear interpolation of the pointwise estimates $\hat f(x_i)$, and we denote it by $f_\text{lin}(x)$. This linear interpolation can be made explicit in 1D. Let us consider the sample points $\{x_i\}$ and let us consider the order statistics $\{x_{(i)}\}$, defined in such a way that $x_{(1)}<x_{(2)}\dots < x_{(N)}$. We now have for a point $x \in [x_{(i)},x_{(i+1)}]$ and with $i=1,,N-1$

\begin{equation}\label{eq:lin_interp}
     f_\text{lin}(x)=\hat f_\text{KDE}(x_{(i)})+\frac{x-x_{(i)}}{x_{(i+1)}-x_{(i)}}\left(\hat f_\text{KDE}(x_{(i+1)})-\hat f_\text{KDE}(x_{(i)})\right).
\end{equation}
Clearly,  it holds that $ f_\text{lin}(x_{(i)})=\hat f_\text{KDE}(x_{(i)})$, since the estimate in the points of the sample does not depend on the interpolation. 
So, the second estimator built upon a linear interpolation is then written as
\be\label{eq:interp2}
\hat f_2(x)=C_2\left(f_\text{lin}(x) - b\frac{\, K(0)}{N h}\right)\,  I_{\mathbb{D}_2}(x),  \quad \text{ with } \mathbb{D}_2=[x_{(1)},x_{(N)}]\,,
\ee
where $\mathbb{D}_2$ is the convex-hull of the sample points. In this case, $\mathbb{D}_2$ is compact and  we have $\hat f_2(x)>0, \forall x \in \mathbb{D}_2$ 
(at least for  kernels with non-compact support). Like before,  we fix $C_2$ by the normalization condition and one gets $\hat f_2(x)\geq 0, \forall x\in \mathbb{R}$ and $\int_\mathbb{R} \, \hat f_2 (x)\, \dif x = 1$. 

Notice that the estimator discussed in the main text is the second one $\hat f_2$.

In this appendix we keep the technicalities to a minimum, postponing more details into \cref{sec:technical}. We  use results from the KDE error analysis, which holds in the case that $f''(x)$ is continuous and there are no boundary bias terms. Under the same requirements, we have that 
\be
C_1,C_2\leq 1+O\left(\frac{x_{(N)}-x_{(1)}}{N h}\right)+O\left(\frac{1}{N}\right)\,.
\label{eq:orderC12}
\ee
This result holds for both estimators \cref{eq:interp1,eq:interp2}, so in the following we will omit the index on the normalization constant $C$ and keep the discussion general.

All the points of the sample belong to both $\mathbb{D}_1$ and $\mathbb{D}_2$, so there is no need to distinguish between the two estimators when talking about the mean square error. We now turn to compute the bias and variance by leaving $C$ unspecified 
and making use of the results from the standard KDE analysis (see e.g.~Ref.~\cite{scott2015multivariate}).

For the bias, we get
\begin{align}
\text{Bias}[\hat f] (x_i)&=\mathbb{E}[\hat f](x_i)- f(x_i)=C  \mathbb{E}[\hat f_\text{KDE}](x_i)-\frac{C}{N h} b K(0)-f(x_i)=\nonumber \\
&=C\left[f(x_i) \int K(w)\dif w -h f'(x_i)\int w K(w)\dif w+\frac{1}{2}h^2 \sigma_K^2 f''(x_i)\right]+\nonumber\\
&\quad - f(x_i)- b\frac{C K(0)}{N h}+O(h^4)=\nonumber\\
&=(C-1)f(x_i)+\frac{C}{2}h^2 \sigma_K^2f''(x_i)- b\frac{C K(0)}{N h}+O(h^4)\,,
\end{align}
where $\sigma_K^2$ is the variance of $K(x)$, $\sigma_K^2\equiv\int x^2 K(x)\dif x$.
For the variance we get
\be
\text{Var}[\hat f](x_i)=C^2 \text{Var}[\hat f_\text{KDE}](x_i)=C^2 \frac{R(K)}{Nh}f(x_i)-C^2 \frac{f^2(x_i)}{N}\,,
\ee
where we defined the $L^2$-norm of a generic function $\phi$ as
\be 
R(\phi)\equiv \int \phi^2(x)\,dx\,.
\ee
Putting everything together we have the MSE on the points of the sample $x_i$: 
\be
\text{MSE}(x_i)=\left[(C-1)f(x_i)+\frac{C}{2}h^2\sigma_K^2f''(x_i)-b \frac{C K(0)}{N h}\right]^2+C^2 \frac{R(K)}{Nh}f(x_i)-C^2\frac{f^2(x_i)}{N}+\textrm{h.o.}
\label{eq:MSEi}
\ee
Notice that it is possible to reduce the bias with an appropriate choice of $C$: the bias in the bulk of the sample is lowered for a value of $C$ slightly larger than 1. We will see in \cref{sec:interp1,sec:interp2} that this holds for our proportionality constants $C_1$ and $C_2$.

We now look at the behavior for $N\rightarrow \infty$ and so we have to specify the order of $C$, see \cref{eq:orderC12}.
If we want to discuss the asymptotic behavior we need to specify how $x_{(N)}-x_{(1)}$ depends on $N$. Both $x_{(1)}$ and $x_{(N)}$ are points of the sample, so they are independent of $h$. Formally, this is a problem of order statistics, more details can be found in \cref{sec:technical}. We parametrize this contribution as $x_{(N)}-x_{(1)}=V(N)$, where  upper bounds can be found under suitable regularity assumptions on $f(x)$.
From \cref{eq:MSEi}, and keeping only the lowest orders in $1/N$ and $h$, we get the asymptotic (averaged) mean square error (AMSE) 
\be\label{eq:AMSE}
\text{AMSE}\leq \frac{1}{N}\sum_{i=1}^N \left[\left(K(0)\frac{V(N)}{N h}f(x_i)+\frac{1}{2}h^2\sigma_K^2f''(x_i)\right)^2+ \frac{R(K)}{Nh}f(x_i)\right]
\ee
We know that the requirements needed for KDE in order to be consistent are $N\rightarrow\infty$, $h\rightarrow 0$ and $N h\rightarrow\infty$. In this case a further requirement is necessary: $V(N)/N h\rightarrow 0$. We can see that all these requirements can be satisfied at the same time as long as $V(N)$ does not grow faster than $N$. Indeed, under the condition that $f(x)$ is smooth enough to have a well-defined mean and variance, we know that $V(N)$ cannot grow faster than $\sqrt{N}$ \cite{gumbel1954}. This is sufficient to ensure that AMSE goes to 0 for $N\rightarrow\infty$.

This expression looks very similar to the one for KDE, except for the volume term with $V(N)$. It is instructive to see what happens if the volume term is indeed not there; in such a case, we recover the usual expressions for the optimal bandwidth $h^*$
and the corresponding optimal AMSE
\begin{align}
h^* &=\left(\frac{R(K)}{\sigma_K^4}\right)^{1/5}N^{-1/5}\left(\frac{\sum_i f(x_i)}{\sum_i (f''(x_i))^2}\right)^{1/5}\\
\text{AMSE}^*&=\frac{5}{4}\left(\sigma_K R(K))^{4/5}\right) \langle f''^2\rangle^{1/5}\langle f\rangle^{4/5}N^{-4/5},
\end{align}
where $\langle\cdot\rangle$ indicates the average over the sample points.
This is very useful in order to understand whether the volume term adds sub-leading terms or slows down the convergence.

Under the condition that $f(x)$ admits mean and variance, we derive that the convergence of the AMSE$^*$ to 0 is always guaranteed and it is always faster than $N^{-2/3}$. This rate of convergence is slightly worse than the optimal rate of convergence of KDE, which is $N^{-4/5}$. By looking at the first two terms of \cref{eq:AMSE}, we can see that the condition for the volume term to be subleading is that the volume should not grow faster than $N^{2/5}$. If this is the behavior of that term, then the AMSE is the same as the one of KDE, and thus AMSE$^*$ goes to zero as $N^{-4/5}$. A family of functions for which this is guaranteed are the functions admitting a moment generating function (see Lemma 5.1 of \cite{van2014probability} and the discussion in \cref{sec:technical}). In this case, $V(N)$ cannot grow faster than $\log N$, so the volume term is subleading and we recover the asymptotic properties of KDE.

In our error analysis we relied on the link with KDE in order to see the consistency of the estimator. This consistency has been shown pointwise, meaning on the points of the sample. An analysis of global errors, such as the convergence of MISE to 0, clearly needs a different discussion for each of the two estimators, but  we can justify it as follows. For $\hat f_1(x)$ the MSE($x_i$) can be extended to every point $x$ of the domain $\mathbb{D}_1$. So the MISE analysis would follow straightforwardly from MSE$(x)$. However, the same does not hold for $\hat f_2(x)$, since the value in a generic $x\in\mathbb{D}_2$ depends on the linear interpolation in \cref{eq:lin_interp}, making the link with KDE hard to implement in the MISE analysis. As a first raw approximation we could consider the trapezoidal integration of MSE($x_i$), since the error is known pointwise. In this approximation the MISE analysis would be similar to the AMSE analysis and consistency is guaranteed under the same conditions.

Let us conclude with a discussion on the consistency of the estimators 
\cref{eq:interp1,eq:interp2} at higher dimensionalities.
More practical information about the behavior at higher $D$ and finite $N$ can be found in \cref{sec:tips}. 
The extension of \cref{eq:AMSE} with a generalized bandwidth matrix $H$ is  straightforward
\be\label{eq:AMSED}
\text{AMSE}\leq \frac{1}{N}\sum_i \left[\left(K(0)\frac{V(N)}{N h^D}f(x_i)+\frac{1}{2}h^2\Sigma(x_i)\right)^2+ \frac{R(K)}{Nh^D}f(x_i)\right]\,,
\ee
where $\Sigma(x_i)\equiv\Tr(A^T A \nabla^2 f(x_i))$ and $A=H/h$.
It is also clear that, in this case, the rate of convergence of the AMSE is determined by the behavior of $V(N)$. As for the 1-dimensional case, the requirement for consistency is that $V(N)$ does not grow faster than $N$. In this case all the conditions $N\rightarrow\infty$, $h\rightarrow 0$ and $N h\rightarrow\infty$ can be satisfied at once,  and they are sufficient to ensure consistency. 
If we also have that $V(N)$ does not grow faster than $N^{2/(D+4)}$, then the volume term is sub-leading and we recover the asymptotic properties of KDE.

\section{Comparison between the two estimators}\label{sec:technical}

In \cref{sec:MSE} we performed the error analysis for our density estimators. This section is dedicated to the discussion of more technical details, in particular about the contributions to the normalization constants $C_1$ and $C_2$. We conclude this appendix with some practical discussion about the relative advantages and disadvantages of the two estimators. The kernels considered in this discussion are the exponential, the normal and several compact ones (Epanechnikov, uniform, triangular, cosine).

\subsection{KDE extension}\label{sec:interp1}

Let us remember our first estimator \cref{eq:interp1}
\be
\hat f_1(x)=C_1\left(\hat f_\text{KDE}(x) - b\frac{\, K(0)}{N h}\right)\, I_{\mathbb{D}_1}(x), \quad \text{ with } \mathbb{D}_1=
\left\{x \left\vert\hat f_\text{KDE}(x)\geq b\frac{K(0)}{Nh}\right.\right\},
\ee
in this case we extended the pointwise estimator to the whole domain by using KDE. What we mean by this is that in all the points $x$ such that $\hat f_\text{KDE}(x)\geq b K(0)/ Nh$ the estimate will be proportional to $\hat f_\text{KDE}(x)- b K(0)/Nh$. Clearly this procedure makes this estimator more similar to KDE with respect to $\hat f_2(x)$. This estimator not only becomes similar to KDE for $N\rightarrow\infty$, but it also reduces to the KDE estimator for $b=0$, differently to what happens to $\hat f_2$.

It is now convenient to introduce a new set of points $\mathbb{D}_2'=[x_a,x_b]$, where we have $x_a=\min x\in \mathbb{D}_1$ and $x_b=\max x\in \mathbb{D}_1$.  We also introduced the set of points $\mathbb{D}_2=[x_{(1)},x_{(N)}]$ in \cref{sec:MSE}. The two sets $\mathbb{D}_2$ and $\mathbb{D}_2'$ have similar names because they look very similar and more often than not they do coincide. In order to understand this we have that surely $x_{(1)},x_{(N)}\in \mathbb{D}_1$, but we also know that the typical length scale of the kernels is $h$, so we can  state that $x_a=x_{(1)}-O(h)$ and $x_b=x_{(N)}+O(h)$. For compact kernels we can even be more precise and say that  $x_{(1)}-h \leq x_a \leq x_{(1)}$ and $x_{(N)}\leq x_b\leq x_{(N)}+h$. $\mathbb{D}_2$ and $\mathbb{D}_2'$ also share the property of being compact.

Let us now discuss the behavior of $C_1$.
By requiring the normalization we have:

\be
C_1 \int_{\mathbb{D}_1}\, \hat f_1(x)=1\quad \rightarrow \quad C_1 \left(\int_{\mathbb{D}_1}\, \hat f_\text{KDE}(x)\dif x- b\frac{K(0)}{Nh} V(\mathbb{D}_1)\right)=1 
\ee
We will employ the property of the KDE estimate: $\int_\mathbb{R}\,\hat f_\text{KDE}(x)\dif x=1$. We will also use the fact that $\mathbb{D}_1\subseteq \mathbb{D}_2'$. We then rewrite the integral as
\be
\int_{\mathbb{D}_1}\, \hat f_\text{KDE}(x)\dif x=1-\int_{\overline{\mathbb{D}}_1}\, \hat f_\text{KDE}(x)\dif x=1-\int_{\overline{\mathbb{D}}_2'}\, \hat f_\text{KDE}(x)\dif x-\int_{\mathbb{D}_2'-\mathbb{D}_1}\, \hat f_\text{KDE}(x)\dif x,
\ee
where  an overline indicates the complement of a set with respect to $\mathbb{R}$: $\overline{\mathbb{D}}\equiv \mathbb{R}\backslash \mathbb{D}$.
Finally we can conclude
\be
C_1^{-1}=1-\int_{\overline{\mathbb{D}}_2'}\, \hat f_\text{KDE}(x)\dif x-\int_{\mathbb{D}_2'-\mathbb{D}_1} \hat f_\text{KDE}(x)\dif x- b\frac{K(0)}{Nh} V(\mathbb{D}_1).
\label{eq:C1inverse}
\ee
Notice that there are three contributions that shift $C_1^{-1}$ from 1 and these contributions are all negative. Since they are small, in the expansion they turn into positive contributions to $C_1$.
We now turn to describe each contribution in \cref{eq:C1inverse} separately.

\begin{itemize}
    \item Boundary term: $\int_{\overline{\mathbb{D}}_2'}\, \hat f_\text{KDE}(x)\dif x$.
    
    This term is of order $1/N$. This can clearly be seen by expressing the kernel explicitly:

    \begin{align}
    \int_{\overline{\mathbb{D}}_2'}\, \hat f_\text{KDE}(x)\dif x&=\int_{-\infty}^{x_a}\, \hat f_\text{KDE}(x)\dif x+\int^{+\infty}_{x_b}\, \hat f_\text{KDE}(x)\dif x=\nn\\
    &=\hat F_\text{KDE}(x_a)+1-\hat F_\text{KDE}(x_b)\leq h( \hat f_\text{KDE}(x_a)+\hat f_\text{KDE}(x_b))\nn\\
    &=2\frac{K(0)}{N}.
    \end{align}
The inequality in the second line is the least obvious step. 
That inequality holds evidently for a compact kernel, while an explicit computation is needed for other kernels, in particular the inequality becomes an equality for the exponential kernel.
    
    \item Discontinuity in the domain: $\int_{\mathbb{D}_2'-\mathbb{D}_1}\, \hat f_\text{KDE}(x)\dif x$.
    
    As easily understood, $\mathbb{D}_1$ can be non-compact and the normalization constant $C_1$ receives contributions because we are setting the estimate to 0 in parts of $\mathbb{R}$ where $\hat f_\text{KDE}$ is not 0. However, by definition, we have that $\hat f_\text{KDE}\leq b K(0)/Nh$ in this situation, this allows us to set an upper bound:
    
    \be
    \int_{\mathbb{D}_2'-\mathbb{D}_1} \, \hat f_\text{KDE}(x)\dif x\leq b \frac{K(0)}{Nh}    \int_{\mathbb{D}_2'-\mathbb{D}_1} \, \dif x = b \frac{K(0)}{Nh} (V(\mathbb{D}_2')-V(\mathbb{D}_1)).
    \ee
    
    Since these volumes depend  on the specific sample and also on $h$, let us put this term together with the next one.
    
    \item Volume term: $b\frac{K(0)}{Nh} V(\mathbb{D}_1$).
    
    We can add this term to the previous one and get
\begin{align}
    \int_{\mathbb{D}_2'-\mathbb{D}_1} \, \hat f_\text{KDE}(x)\dif x+ b \frac{K(0)}{Nh} V(\mathbb{D}_1) &\leq b \frac{K(0)}{Nh} (V(\mathbb{D}_2')-V(\mathbb{D}_1))+b\frac{K(0)}{Nh} V(\mathbb{D}_1)\nn\\ &=b\frac{K(0)}{Nh} V(\mathbb{D}_2').
\end{align}
    This is the mentioned term of order $V(N)/Nh$. We have that $V(\mathbb{D}_2')=V(\mathbb{D}_2)+O(h)=x_{(N)}-x_{(1)}+O(h)$. So, in the end we have
    \begin{align}
    \int_{\mathbb{D}_2'-\mathbb{D}_1} \, \hat f_\text{KDE}(x)\dif x+ b \frac{K(0)}{Nh} V(\mathbb{D}_1) &\leq b \frac{K(0)}{Nh}  V(\mathbb{D}_2')\nn\\ 
    &= b \frac{K(0)}{Nh} (x_{(N)}-x_{(1)})+O\left(\frac{1}{N}\right).
    \end{align}

\end{itemize}
This last contribution $K(0) (x_{(N)}-x_{(1)})/Nh$ will be the leading one and hence the one relevant for error analysis. 
So we now need to study  $x_{(1)}$ and $x_{(N)}$, as they control the volume of $\mathbb{D}_2'$. 
These two points are the smallest and largest point of the sample. Because of this,  they depend on the $f(x)$ we are estimating and on $N$, but not on $h$. From order statistics we know the expressions for their expected values as a function of $N$ and $f(x)$ \cite{casella2002statistical}. 

We use two bounds on $x_{(1)}$ and $x_{(N)}$ deriving from properties of $f(x)$. A first loose bound comes from the requirement that $f(x)$ admits a mean $\mu$ and a variance $\sigma^2$ \cite{gumbel1954}. In this case we know that $\langle x_{(1)} \rangle \geq \mu -\sigma \sqrt{N}$ and $\langle x_{(N)} \rangle \leq \mu +\sigma \sqrt{N}$. This requirement is very loose, by being more strict and asking that the function $f(x)$ admits a moment generating function (a requirement equivalent to admitting a bilateral Laplace transform or admitting moments of every order), we can also derive that $x_{(N)}$ cannot grow faster with $N$ than $\log N$. This can be understood by looking at Lemma 5.1 of \cite{van2014probability}. In particular, in the proof we find the inequality
\be\label{eq:max_gaussians}
\langle x_{(N)}\rangle\leq \inf_{s>0} \frac{\log N + \log (m(s))}{s}, 
\ee
where $m(s)$ is the moment generating function, which is a monotonically increasing function of $s$ and taking values between $1$ and $+\infty$. This means that for any $N$ we can find $\bar{s}$ such that $m(\bar s)=N$. In particular, since $m(s)$ is increasing with $s$, we have that increasing $N$ will also increase $\bar{s}$. This last observation tells us that increasing $N$, the denominator of \cref{eq:max_gaussians} will also increase. Putting everything together we conclude that $x_{(N)}-x_{(1)}$ cannot grow faster than $\log N$ for functions admitting a moment generating function.

Let us summarize our results. The quantity $C_1^{-1}$ is equal to 1 except for three negative contributions. A boundary term of order $1/N$ and a term which is less than $V(N)/N h$ in absolute value. Depending on the properties of $f(x)$ we can set some upper bounds on $V(N)$: if $f(x)$ admits first and second central moments, then $V(N)$ cannot increase faster than $N^{1/2}$. If $f(x)$ admits all moments, then we know that it cannot increase faster than $\log N$.

\subsection{Linear interpolation}\label{sec:interp2}

Let us now recall the second estimator \cref{eq:interp2}, obtained from a linear interpolation, is
\be
\hat f_2(x)=C_2\left( f_\text{lin}(x) - b\frac{\, K(0)}{N h}\right)\,  I_{\mathbb{D}_2}(x),  \quad \text{ with } \mathbb{D}_2=[x_{(1)},x_{(N)}].
\ee
One might wonder why do we bother constructing two estimators, since already the first one is consistent. The reasons for this will be made explicit in \cref{sec:tips}, but let us mention for now that this second estimator behaves differently for finite $N$ as we will shortly see. 

Also in this case we fix the constant $C_2$ by requiring the normalization of $\hat f_2(x)$
\be
C_2 \int_{\mathbb{D}_2}\, \hat f_2(x)=1\quad \rightarrow \quad C_2 \left(\int_{\mathbb{D}_2}\, f_\text{lin}(x)\dif x- b\frac{K(0)}{Nh} V(\mathbb{D}_2)\right)=1 
\ee
First, let us point out that the integral of the linear interpolation is nothing other than the integral calculated with the trapezoidal rule over a non-uniform grid. This grid will have nodes at $x_{(k)}$ and spacings $h_k\equiv x_{(k+1)}-x_{(k)}$. Let us use here the properties of KDE as well
\begin{align}
\int_{\mathbb{D}_2}\,  f_\text{lin}(x)\dif x
&=\int_{\mathbb{D}_2} \left( f_\text{lin}(x)-\hat f_\text{KDE}(x)\right)\,\dif x\,+\int_{\mathbb{D}_2} \hat f_\text{KDE}(x)\dif x\nn\\
&=1+ \delta_\text{trapz}-\int_{\overline{\mathbb{D}}_2}\, \hat f_\text{KDE}(x)\dif x,
\end{align}
where we have defined $\delta_\text{trapz}\equiv\int_{\mathbb{D}_2} \left( f_\text{lin}(x)-\hat f_\text{KDE}(x)\right)\dif x$. This term can be a priori either positive or negative.

Let us make the contributions to $C_2$ explicit
\be
C_2^{-1}=1-\int_{\overline{\mathbb{D}}_2}\, \hat f_\text{KDE}(x)\dif x- b\frac{K(0)}{Nh} V(\mathbb{D}_2)+\delta_\text{trapz}.
\label{eq:C2inverse}
\ee
This expression is very similar to the one derived for $C_1$. The difference lies in the addition of the third term $\delta_\text{trapz}$. We will see that this contribution is usually positive, thus partially compensating the other two negative contributions.

In the treatment of the contributions, in order to be as general as possible, we will consider two portions of $\mathbb{D}_2$: one where we have $h_k\leq h$ asymptotically and the other where we have $h_k>h$. Notice that the average value of $h_k$ is given by $V(\mathbb{D}_2)/N$, this means that there will always be a portion of $\mathbb{D}_2$ where $h_k\leq h$. 

We now turn to describe each contribution in \cref{eq:C2inverse} separately.

\begin{itemize}
    \item Boundary term: $\int_{\overline{\mathbb{D}}_2}\, \hat f_\text{KDE}(x)\dif x=\int_{-\infty}^{x_{(1)}}\, \hat f_\text{KDE}(x)\dif x+\int^{+\infty}_{x_{(N)}}\, \hat f_\text{KDE}(x)\dif x$.
    
    This term is of order $1/N$ under the assumptions we are considering. Let us express this more explicitly
\begin{align}
    \int_{-\infty}^{x_{(1)}}\, \hat f_\text{KDE}(x)\dif x+\int^{+\infty}_{x_{(N)}}\, \hat f_\text{KDE}(x)\dif x
    &=\hat F_\text{KDE}(x_{(1)})+1-\hat F_\text{KDE}(x_{(N)})\nn\\
    &\leq h( \hat f_\text{KDE}(x_{(1)})+f_\text{KDE}(x_{(N)})).
\end{align}
    
    The inequality comes from the same reasoning carried out in the previous section. If we have $h_1, h_{N-1}\gtrsim h$, then the extremal points of the sample are isolated and we have $\hat f_\text{KDE}(x_{(N)})=(Nh)^{-1}\sum K(x_{(N)}-x_i)\simeq (Nh)^{-1}K(0)$ and we recover the order $1/N$. This approximated equality becomes  an equality for compact kernels and $h_1, h_{N-1}> h$. The same holds for $x_{(1)}$.
    Terms of order $h$ could be present, if $f(x_{(1)})$ or $f(x_{(N)})$ would not go to 0 for increasing $N$. However, this would imply that KDE has problems with boundary bias, a condition we need to avoid in order to carry out the error analysis. Ways to treat problems on the boundary are discussed in \cref{sec:tips}.

    \item Volume term: $b\frac{K(0)}{Nh} V(\mathbb{D}_2$).
    
    The discussion of this term is identical to the one carried out in the previous section and the same upper bounds on its behavior hold.
    
    \item $\delta_\text{trapz}: \int_{\mathbb{D}_2} \left( f_\text{lin}(x)-\hat f_\text{KDE}(x)\right)\dif x$
    
    Here it is very useful to distinguish between the two regimes of $h_k$. We will show that this term is (asymptotically) less than the volume term with $b=1$ in absolute value. In order to easily compare, let us split the volume contribution and this contribution over the intervals between each sample point
\begin{align}
\delta_\text{volume}&=\frac{K(0)}{Nh} V(\mathbb{D}_2)=\frac{K(0)}{Nh} \sum_{k=1}^{N-1} h_k=\sum_{k=1}^{N-1} h_k \delta_\text{volume}(k)\\
\delta_\text{trapz}&=\sum_{k=1}^{N-1} \frac{h_k}{N h}\sum_i\left[ K\left(\frac{x_{(k)}+h_k-x_i}{h}\right)
    +K\left(\frac{x_{(k)}-x_i}{h}\right)\right. \nn\\
    &\left. \hspace{3cm} -\frac{\int_{x_{(k)}}^{x_{(k)}+h_k}K(\frac{x-x_i}{h})\dif x}{h_k} \right]
    =\sum_{k=1}^{N-1} h_k \delta_\text{trapz}(k)
\end{align}
In the first regime we have $h_k/h<1$, so we can expand around small $h_k$ getting (this is nothing else that the error for the trapezoidal rule):
    
    \be
    \delta_\text{trapz}(k)=\frac{h_k^2}{12 N h^3 }\sum_i K''\left(\frac{x_{(k)}-x_i}{h}\right)\,.
    \ee
    
    Now by comparing interval by interval we would like to show that:
    
    \be
    \delta_\text{volume}(k)=\frac{K(0)}{N h}\geq \frac{h_k^2}{12 N h^3}\sum_i K''\left(\frac{x_{(k)}-x_i}{h}\right).
    \ee
    In order to show this we first consider that only a finite number of terms is contributing, namely all $x_i$ such that $x_{(k)}-x_i\lesssim h$. The number of such points is of the order of $h/h_k$. Secondly, we have that for all the kernels we considered $K''((x_{(k)}-x_i)/h)\leq |K''(0)|\leq \gamma K(0)$. Then putting all the inequalities together we have
    \be
    \frac{h_k^2}{12 N h^3}\sum_i K''\left(\frac{x_{(k)}-x_i}{h}\right)\lesssim \frac{h_k^2}{12 N h^3} \gamma K(0) \frac{h}{h_k}= \frac{\gamma}{12}\frac{h_k}{h} \frac{K(0)}{ N h}.
    \ee
    
    This term is indeed smaller than ${K(0)}/({N h})$ for $h_k\leq h$. Intuitively and from numerical simulations, we gather that the contribution to $\delta_\text{trapz}$ in this regime will be much smaller than the one when $h_k> h$.
    
    In the second regime we have isolated points, this means that we have $f_{x_{(k)}}\simeq f_{x_{(k+1)}}\simeq K(0)/Nh$ (this equality is exact for compact kernels). In this case we cannot expand around small $h_k$, but it is easy to evaluate $\delta_\text{trapz}$ explicitly
    \be
    \delta_\text{trapz}(k)=\frac{1}{h_k}\left(\int_{x_{(k)}}^{x_{(k)}+h_k} \left( f_\text{lin}(x)-\hat f_\text{KDE}(x)\right)\dif x\right)\simeq\frac{K(0)}{N h}- \frac{K(0)}{N h_k}.
    \ee
    This is equal in absolute value to the volume term in the asymptotic limit and smaller in the other cases, but it is always positive. Since this is in general larger than the previous contribution, we can safely state that in general $\delta_\text{trapz}$ is positive.
    
    This description of $\delta_\text{trapz}$ would break down if there were strong discontinuities in the spacings $h_k$. These can be due to two effects: discontinuities in $f(x)$, which can be avoided by requiring a continuous $f(x)$. Or they can be due to statistical fluctuations, these are however not supposed to be present in the asymptotic limit $N\rightarrow \infty$.
    
    We need to make a further comment for what concerns compact kernels. In the case $h_k\leq h$ we have three subcases (let us assume $x_i<x_{(k)}$ for simplicity): $x_i+h<x_{(k)}$ and in this case there is no contribution to $\delta_\text{trapz} (k)$. Then we can have $x_i+h>x_{(k)}+h_k$ and the contribution is the same described before. Finally, we have a contribution if $x_{(k)}<x_i+h<x_{(k)}+h_k$. In this case the number of points contributing will be order 1 and the scaling of the contribution will depend on the specific kernel. More explicitly we will have contributions of the form:
    
    \be \label{eq:compact}
    \delta_\text{trapz}(k)=\frac{1}{N h}\sum_i\left( K\left(\frac{x_{(k)}-x_i}{h}\right) -\frac{1}{h_k}\int_{x_{(k)}-x_i}^h K\left(\frac{x}{h}\right)\dif x\right)\leq \frac{C}{N h_k} \left(\frac{h_k}{h}\right)^\alpha a^\beta,
    \ee
    
here $C$ is an order 1 term with subleading dependencies on $h_k/h$ and $a$. $a$ is defined as $a\equiv (x_i+h-x_{(k)})/h_k$ and so $a\in [0,1]$. $\alpha$ and $\beta$ are kernel dependent, but for all compact kernels (except the uniform) we have $\alpha\geq 2$ and $\beta\geq 1$, so this additional term gives no problems. For the uniform kernel we have $\alpha=1$, but we also have positive contributions for $a<1//2$ and negative for $a>1/2$. So this term is not problematic if there are no strong discontinuities in $h_k$, a condition we had already imposed to evaluate the trapezoid contribution.

\begin{table}[t]
    \centering
    \begin{tabular}{c|c|c|c|c|c}
         Kernel & $K(u)$ & support & $\alpha$ & $\beta$ & $\gamma$ \\
         \hline
         \hline
         Gaussian & $\frac{1}{\sqrt{2\pi}}e^{-u^2/2}$& $\mathbb{R}$& - & - & 1 \\
         Exponential & $\frac{1}{2}e^{-|u|}$& $\mathbb{R}$ & - & - & 1 \\
         Uniform & $1/2$ & $|u|\leq1$&1 & 0 & 0 \\
         Triangular/Linear &$1-|u|$ &$|u|\leq1$& 2 & 1 & 0\\
         Epanechnikov &$\frac{3}{4}(1-u^2)$& $|u|\leq1$& 2 & 1 & 2 \\
         Cosine & $\frac{\pi}{4}\cos\left(\frac{\pi}{2} u\right)$& $|u|\leq1$ & 2 & 1 & $\pi^2/4$\\
         Biweight & $\frac{15}{16}(1-u^2)^2$& $|u|\leq1$ & 3 & 2 & 4\\
         Triweight & $\frac{35}{32}(1-u^2)^3$& $|u|\leq1$ & 4 & 3 & 6\\

    \end{tabular}
    \caption{Values of $\gamma\equiv |K''(0)|/K(0)$, $\alpha,\beta$. $\alpha$ and $\beta$ are defined in \cref{eq:compact} and apply only to compact kernels. We reported the expression for the Kernel functions as implemented in KDE. We usually expressed the kernel as a function of the distance, in which case we just need to take $d=h|u|$ in 1D. Biweight and triweight have not been implemented in the software we provide, but they would give no issues with the performance as can be understood from the values of $\alpha,\beta,\gamma$.}
    \label{tab:kernels}
\end{table}

\end{itemize}
Let us summarize our results. The quantity $C_2^{-1}$ is equal to 1 except for three contributions: two always negative, one usually positive. The volume term and the trapezoid term are both bounded in absolute value by something of the order of $V(N)/N h$. Depending on the properties of $f(x)$ we can set some upper bounds on $V(N)$: if $f(x)$ admits first and second central moments, then $V(N)$ cannot increase faster than $N^{1/2}$. If $f(x)$ admits all moments, then we know that it cannot increase faster than $\log N$. Then there is a boundary term which is of order ${1}/{N}$, under the assumption that there is no boundary bias. 

\subsection{Practical tips} \label{sec:tips}

The previous part of this section was dedicated to a comparison between the asymptotic properties of the two estimators \cref{eq:interp1,eq:interp2}. The discussion there was fairly theoretical and we now turn to more practical distinctions between the two. We also conclude by mentioning how to treat boundary bias terms, a problem afflicting our estimators and KDE alike.

\subsubsection*{Large N, small N}

Asymptotic properties for $N\rightarrow\infty$ are important to show consistency, but in practice we always work with finite $N$. In the following we will often refer to properties that hold for large or small $N$. These concepts of large and small $N$ are loose, by large $N$ we mean a situation where the domain is densely populated and we are then close to the asymptotic description. On the contrary, small $N$ indicates that the population is rather sparse. It should be clear that these concepts depend on the dimensionality and the distribution we are studying: $N=1000$ is large if we are studying a unimodal distribution in 1D, but it is rather small if we want to study a multimodal distribution in 6D.

\subsubsection*{Optimization}

Under the conditions specified in \cref{sec:MSE} we have that our estimators and KDE share the same asymptotic properties. From this and from \cref{eq:interp0} we understand that for large N our estimators and KDE will be similar. This helps us for optimization. As a matter of fact, since the optimization of KDE is widely discussed, we can rely on finding the optimal bandwidth for KDE and then using that for our estimators. This is true for both theoretical rule-of-thumbs and data-based optimization. Indeed, this is what we used in \cref{fig:comparison3}, where we used a 5-fold cross validation to determine the optimal bandwidth of KDE and then used the same value for our estimators. For these plots we have checked that a 10-fold cross validation or the leave-one-out cross validation would have yield similar results. Of course, this approximation holds better the larger N is. 

Relying on KDE optimization is however not the only option. For both $\hat f_1$ and $\hat f_2$ we can minimize the negative log-likelihood before rescaling with $C_1$ and $C_2$. This is equivalent to considering:
\be \label{eq:loss1}
\text{Loss}(h)=-\sum_i \log \left(\hat f_\text{KDE}(x_i) - \frac{\, K(0)}{N h}\right).
\ee
This loss function shows a clear minimum. We can realize that this loss function works, since it is equal to the leave-one-out log-likelihood function, apart from a constant term independent of $h$. Optimization based on this loss function is very fast with the software provided, since there is an handle that can be used if we do not want to normalize.

The negative log-likelihood of the KDE estimate does not work well as a loss function, this is because there is a trivial minimum for $h=0$, where the estimate becomes a series of delta functions. Similarly, it does not work for $\hat f_1$, even for $b=1$. The situation is different for $\hat f_2$, whose negative log-likelihood
\be\label{eq:loss2}
\text{Loss}(h)=-\sum_i \log (\hat f_2(x_i)).
\ee
shows a single non-trivial minimum for $b=1$. A minimum is usually found also for $0\leq b<1$. This loss function is specific to our estimator and is in some sense ``new", so let us comment a little more on it. Furthermore, it is also the one used in the main text, both for the linear interpolator and the nearest neighbor interpolator. As all loss functions, there are situations for which it does not provide the best possible value of bandwidth. In our simulations we have found that a very good value of the bandwidth is found for $D\geq 2$ for any kernel, while non-optimal values are found at $D=1$ for kernels that are not the Gaussian one. The optimization works better for larger values of $b$. In the main text we worked in the situation $b=1$ and Gaussian Kernel, so everything was perfectly under control.
The value of $h$ found here is in general different from the one derived from \cref{eq:loss1}, since $C_2$ depends on $h$. Since \cref{eq:loss2} requires a longer computation with respect to \cref{eq:loss1} we need to have some advantage in doing so. Indeed, it is clear that \cref{eq:loss2} encodes more information, as \cref{eq:loss1} is calculated before rescaling and does not even distinguish $\hat f_1$ and $\hat f_2$. Also $C_1$ depends on $h$, so we might think that $\text{Loss}(h)=-\sum_i \log (\hat f_1(x_i))$ works well for optimizing $\hat f_1$. That is usually not the case and this can be understood by looking at the contributions to $C_1$ and $C_2$. $C_1$ receives a positive contribution proportional to $1/Nh$, while the dependence of $C_2$ is less trivial, since there are two terms of similar order compensating each other. This implies that $C_1$ (and in turn $\hat f_1$) has a strong dependence on $h$, while the same is not true for $C_2$. This is what ultimately leads to the stability of the loss function in \cref{eq:loss2} and to the fact that the performance of $\hat f_1$ depends on the bandwidth much more than the one of $\hat f_2$.

This optimization procedure gives reliable results and it is the one that relies the least on KDE results, thus being the most self consistent. It has unfortunately the disadvantage of being lengthy to compute, since we are normalizing for each value of $h$. The loss function in \cref{eq:loss2} is the one used in \cref{sec:exp_de}.

\subsubsection*{Numerical comparison between the estimators}

\begin{figure}[t]
    \centering
    \includegraphics[width=0.46\textwidth]{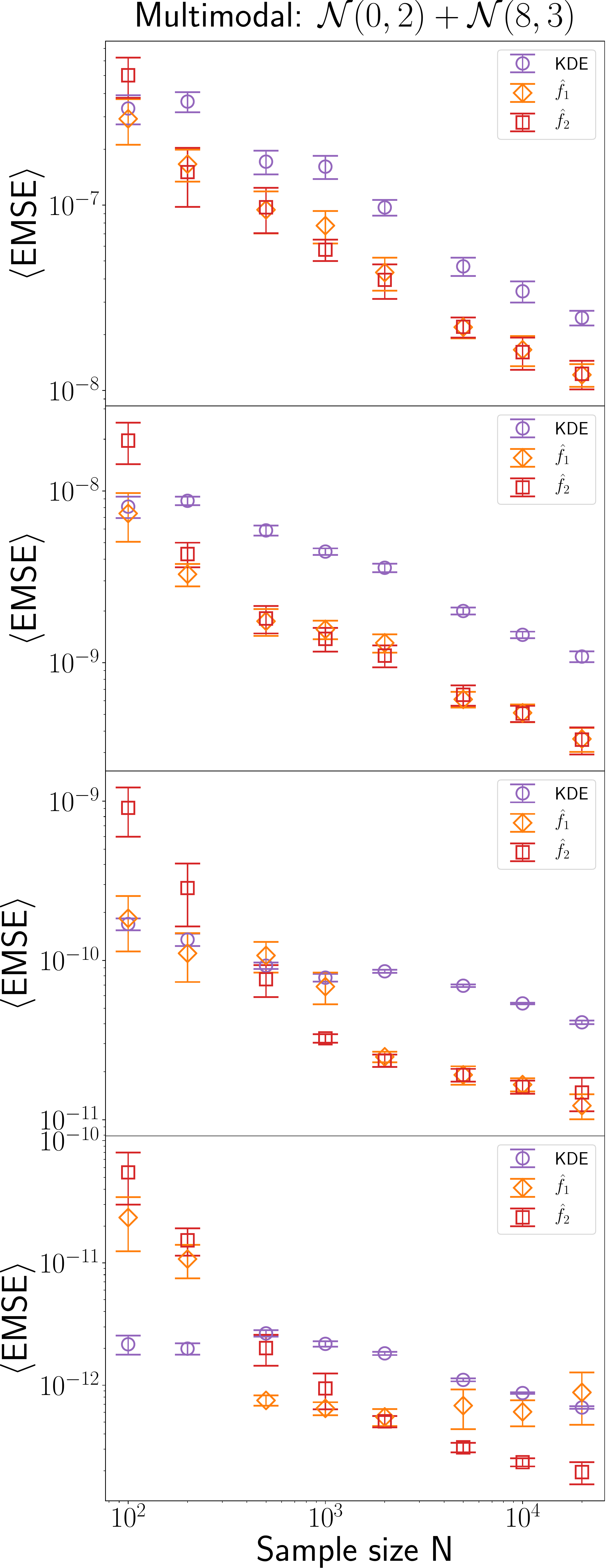}\hspace{8mm}
    \includegraphics[width=0.46\textwidth]{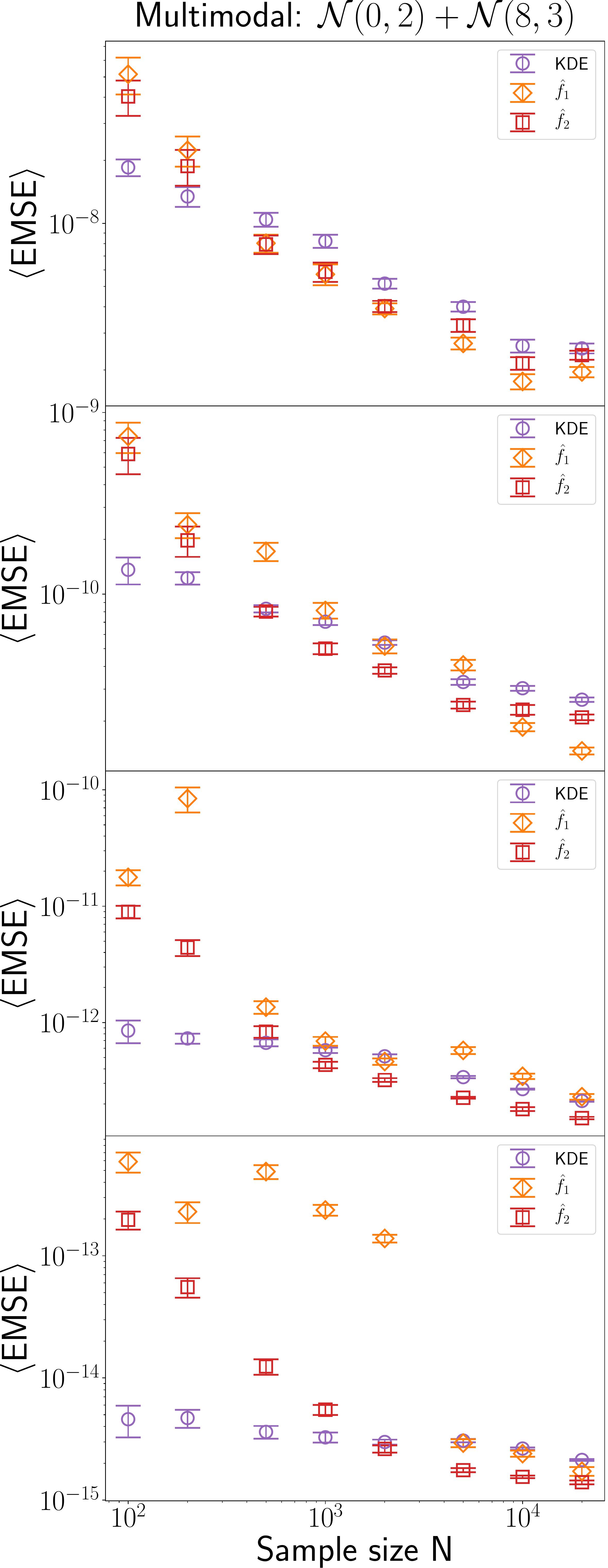}
    \caption{EMSE for KDE and our two estimators for different values of D and with increasing sample size. For all estimators the bandwidth used is the same, chosen by performing 5-fold cross-validation on KDE. The error bars indicate the $2\sigma$ statistical uncertainty evaluated with $R=9$ sample realizations. \emph{Left panel:} Unimodal case: $\chi^2(5)$. \emph{Right panel:} Multimodal case: $\mathcal{N}(0,2)+\mathcal{N}(8,3)$}    \label{fig:comparison3}
\end{figure}

It is useful to directly compare the results from KDE and of two estimators on some explicit cases. This is reported in  \cref{fig:comparison3}.  In order to compare we have used the same optimal bandwidth for each estimator, this bandwidth was derived with a 5-fold cross validation, the values of bandwidth used were evenly spaced in the linear interval $[0.05,1]$. The reason we do this is to highlight the different behaviors of the estimators, rather than the effects of optimization. We report the average mean square error as a measure of the error at different dimensionalities and for different sample sizes. The mean values and the error bars are derived by making $R=9$ sample realizations in the same way we did for \cref{sec:exp_de}. Let us look at what we can say.

First of all, there is no clear winner, meaning that depending on the size of the sample and the dimensionality we get different hierarchies between the estimators. However, some common trend is exhibited.

$\hat f_2$ is the most regular one: bad for small samples and increasingly good as we enlarge the sample. The errors for small samples are due to the poor reconstruction of the interpolator, changing the bandwidth does not improve the result much. $\hat f_2$ is less-sensitive to the $h$ selection with respect to $\hat f_1$, this also explains part of its regularity. 

$\hat f_1$ on the other hand is far less regular, this can be understood because there can be a large positive contribution to $C_1$ as discussed in the previous section. This contribution is inversely proportional to $N h$, so it worsens things for small $h$ and small $N$. This implies that the estimator is well-behaved at large $N$, but not at small $N$. From the plots it is clear that the estimator is most irregular for $D\geq 5$. It is anyway possible to check a posteriori if we are in the well-behaved case or not by looking at $C_1$. When we perform the density estimate we can ask the software to return the value of $C_1$ to us, so we can see how much it differs from 1 and how worried we should be. Differently from the case of $\hat f_2$, here the value of $h$ enters critically in the goodness of the estimate. This strong dependence on $h$ is also perceived by looking at the error bars: the same value of bandwidth can be perfect for a sample and not-so-good for a different sample, even if $D$ and $N$ are the same. Changing (usually enlarging) the bandwidth at small $N$ could improve significantly the final result, as this is most sensitive to the bandwidth. Of course, some different way of optimizing should be introduced. As a practical tip, the optimization procedure that would lead to the largest $h$ is the safest choice for optimizing this estimator. By safest we mean that it is the one that avoids the irregular behavior the most, but it is not necessarily the optimal one.

\subsubsection*{Boundary bias}

A well known problem of KDE is the boundary bias and our estimators can suffer from the same problem, fortunately also many of the solutions can be used also for our estimators. 
There are two cases of interest for us. In the first case, a single boundary point is present, then we can always reduce to the case where $f(x)$ has support on $[0,+\infty)$. In the second case we have two boundary points and the function has support on a compact interval, which can be sent to the unit interval $[0,1]$.

If there is a single boundary, then a simple solution is doubling the points \cite{cline1991kernel}. A generalized reflection which further improves the reduction of the bias when $f'(0)\neq 0$ is proposed in \cite{10.2307/25046198}.

A general solution which applies to both cases is transforming the variable $x$ in such a way that whichever support $f(x)$ has, the new $y=g(x)$ will take positive values on $\mathbb{R}$ \cite{10.2307/2346189,geenens2014probit}. For a discussion on the consistency of the estimator with this change of variable see for instance \cite{geenens2014probit}.

\bibliographystyle{JHEP}
\bibliography{bibliography}

\end{document}